\documentclass[twocolumn,dvipdfmx]{jpsj3}
\usepackage{txfonts}
\usepackage{bm}
\usepackage{amsmath,amssymb}
\usepackage{float}
\usepackage{ascmac}
\usepackage{subfigure}
\usepackage{verbatim}
\usepackage{wrapfig}
\usepackage{makeidx}
\usepackage{layout}
\usepackage{braket}

\newcommand{\cc}{\mathrm{c}}
\newcommand{\n}{\mathrm{n}}
\newcommand{\BdG}{\mathrm{BdG}} 

\newcommand{\ii}{i}
\newcommand{\F}{\mathrm{F}} 
\newcommand{\trans}{{}^{\mathrm{t}}}
\newcommand{\bb}{\mathrm{b}} 
\newcommand{\dint}{\displaystyle\int}
\title{Impurity Effects on Bound States in Vortex Core of Topological s-Wave Superconductor}

\author{Yusuke Masaki$^{1}$\thanks{E-mail: masaki@vortex.c.u-tokyo.ac.jp} and Yusuke Kato$^{2}$}
\inst{$^1$Department of Physics, The University of Tokyo, Tokyo 113-0033, Japan \\
$^2$Department of Basic Science, The University of Tokyo, Tokyo 153-8902, Japan
} 

\abst{
We study the impurity effects on the Caroli-de Gennes-Matricon (CdGM) states, particularly on the level spacings in a vortex core in a topological s-wave superconductor (SC) by two methods, numerical and analytical.
The topological s-wave SC belongs to the same class as a chiral p-wave SC, and thus there are two inequivalent vortices in terms of any symmetry operation.
We take into account this inequivalence and numerically calculate the scattering rates based on an improved version of the Kopnin-Kravtsov (iKK) scheme, which enables us to treat the discrete levels in the presence of white-noise disorder.
We also construct the Andreev equation for the topological s-wave SC and obtain the Andreev bound states analytically. We use a correspondence between the wave functions for 
the Bogoliubov-de Gennes equation and the Andreev equation in the iKK scheme and deduce the formula of scattering rates described by the wave function for the Andreev equation. With this formula, we discuss the origin of impurity scattering rates for CdGM states of the topological s-wave SC and the dependence on the types of vortex related to the inequivalence.
}

\begin{document}
\maketitle

\section{Introduction}
An implementation of topological quantum computation (TQC) with low decoherence is a promising application of topological superconductors (TSCs)~\cite{PhysRevLett.86.268,RevModPhys.80.1083}. 
In addition to this fascinating application, 
recent intensive studies of TSCs have been motivated by fundamental interest in the appearance of the Majorana fermion in condensed matter physics~\cite{RevModPhys.82.3045,RevModPhys.83.1057}.
The Majorana fermion exists as a topologically protected zero-energy bound state near the edge or around the topological defects in TSCs 
and this zero-energy state obeys a non-Abelian statistics.
The implementation of TQC is realized via the braiding operation among the degenerated Majorana zero-energy states~\cite{PhysRevLett.86.268, alicea2011non}. 
It is necessary to perform the braiding operation adiabatically in order to avoid nonadiabatic transitions of zero-energy states. 
The typical operation time should be longer than the time given by $\hbar$ divided by the level spacing between the Majorana state and the first excited state 
(we call this condition the ``adiabatic condition''). 
It is thus important that the level spacing, i.e., the energy of the first excited state, is stable against disorder as well as the zero-energy state also 
in terms of the signal strength of measurements~\cite{PhysRevB.82.174506,PhysRevB.82.020509}. 
In this work, we focus on the Majorana state 
in a vortex core and investigate the impurity effects on the bound states in a vortex core called Caroli-de Gennes-Matricon (CdGM) states~\cite{Caroli1964}.

A chiral p-wave SC~\cite{RevModPhys.59.533,volovik1999fermion,PhysRevB.61.10267,PhysRevLett.86.268} is a typical two-dimensional TSC that belongs to the Bogoliubov-de Gennes (BdG) class with broken time-reversal symmetry, called ``class D''. 
This class is one of ten symmetry classes obtained by Altland and Zirnbauer on the basis of random matrix theory~\cite{PhysRevB.55.1142}, 
and all possible TSCs have been classified into these classes by Schnyder {\it et al.}~\cite{PhysRevB.78.195125, ryu2010topological} if we do not take into account the crystal symmetry. A TSC in class D has two inequivalent single vortices in terms of symmetry operations: the inequivalence is characterized by the relative sign of chirality and vorticity. 
There is a remarkable difference between the two inequivalent vortices, 
according to an earlier work concerning impurity effects on low-energy states localized in a vortex core 
in a chiral p-wave SC by means of quasiclassical theory~\cite{Kato2000,JPSJ.71.1721}. 
The scattering rates of the localized states are almost the same as those in the normal states when the relative sign is positive. 
On the other hand, 
the scattering rates are vanishingly small when the sign is negative. 
This is a consequence of coherence effects and rotational symmetry. 
However, Sr$_{2}$RuO$_{4}$, a candidate for a chiral p-wave SC, 
has a point group symmetry $C_{4}$ in the crystal structure~\cite{MaenoKittakaNomuraYonezawaIshida2012} 
and we could not expect to observe the effect of rotational symmetry.
Other possible candidates for class D TSCs are, for example, the $\nu =5/2$ fractional quantum Hall state~\cite{Moore1991,PhysRevB.61.10267} and some engineered TSCs such as 
the surface of a topological insulator with an s-wave pair potential~\cite{PhysRevLett.100.096407} 
and a semiconductor heterostructure with an s-wave SC and a ferromagnet~\cite{PhysRevLett.104.040502}. 
The latter engineered TSC, 
which is described as a two-dimensional electron gas (2DEG) with Rashba-type spin orbit coupling (SOC), Zeeman coupling, and an s-wave pair potential and we call the topological s-wave SC, 
has rotational symmetry derived from the 2DEG realized in the semiconductor heterostructure. 
We expect that the impurity effects dependent on the symmetry will be observed more clearly in this system.

There are several ways of treating the impurity effects on localized states in a vortex core. 
The impurity strength $\Gamma_{\n}$ should be small such that the low energy spectra are discrete, 
since our motivation is to discuss the level spacing in the presence of impurities.
In this case, we cannot use the quasiclassical theory, in which the spectra are treated as continuous~\cite{Eilenberger1968,larkin1969quasiclassical}. 
Moreover, the Born parameter should be vanishingly small.
In this case, there are many but weak scatterers, and the (self-consistent) Born approximation is valid. 
In a typical unconventional superconductor, 
the density of states (DoS) is formed around the Fermi energy if the Born parameter is not small~\cite{RevModPhys.78.373}. 
Even in an s-wave SC, the impurities, the Born parameter of which is not small, lead to the level mixing 
and the Landau-Zener transition~\cite{PhysRevB.57.5457,PhysRevB.59.12021,PhysRevB.60.14597}.
Considering the above discussion, 
we can calculate the impurity effects using the Kopnin-Kravtsov scheme~\cite{kopnin1976conductivity}, 
in which Green's function is used for the CdGM mode with impurity self energy [self-consistent Born approximation (SCBA)] while keeping the levels discrete. 
We have improved their scheme in terms of the coherence factor and applicability to various types of SC in addition to an s-wave SC and call this scheme the improved Kopnin-Kravtsov (iKK) scheme~\cite{masakikato}.

The aim of this work is to understand the impurity effects on the level spacing related to the adiabatic condition and the physical picture of these impurity effects.
To address these issues, we first numerically calculate the scattering rates of the vortex core states based on the iKK scheme and 
evaluate the property of the minigap for a topological s-wave SC in the presence of impurities, in terms of two inequivalent vortices. 
In this case, the sign of chirality related to the chiral edge mode is determined by the sign of the Zeeman coupling~\cite{Wu2012}.
We find that the obtained numerical results are more complicated than those for the chiral p-wave SC, which have been understood only by considering the type of vortex.
We then make use of the Andreev approximation~\cite{andreev1964thermal,volovik1993vortex,Tewari2010219} in order to understand the physical picture of these results
and the origin of the scattering rates. The physical picture can be successfully understood by considering the combinations of two types of vortex for the chiral p-wave SC.
Although the Andreev approximation is a type of quasiclassical theory, 
we can use it to obtain the physical picture  after numerical calculation based on the quantum theory.
We take $\hbar = 1$ throughout this paper.

\section{Model and Method} \label{sec-model-method}
We consider the two-dimensional superconducting system described by the following Hamiltonian~\cite{PhysRevLett.100.096407}: 
\begin{align}
&H = \int d\bm{r} \vec{\underline{\psi}}^{\dagger}(\bm{r})\hat{H}_{0}\vec{\underline{\psi}}^{}(\bm{r})  
+ \int d\bm{r} \left[\Delta(\bm{r}) \psi_{\uparrow}^{\dagger}(\bm{r})\psi_{\downarrow}^{\dagger}(\bm{r}) + \mathrm{h.c.}\right],  \label{hamiltonian} \\
&\hat{H}_{0}= \left[\dfrac{\bm{p}^{2}}{2m}-\mu+ \alpha (\bm{p}\times \hat{\bm{\sigma}})_{z}+V_{z}\hat{\sigma}_{z}\right],  
\ \vec{\underline{\psi}}^{\dagger}(\bm{r}) = (\psi_{\uparrow}^{\dagger}(\bm{r}),\psi_{\downarrow}^{\dagger}(\bm{r})),
\end{align}
where $\mu, \alpha$, and $V_{z}$ denote 
the chemical potential, the strength of Rashba SOC, and the Zeeman coupling, respectively. 
$\hat{\sigma}_{x,y,z}$ are 2 by 2 Pauli matrices and the symbol $\vec{\underline{\ }}$ denotes a two-component vector.
The second term on the right-hand side of Eq. \eqref{hamiltonian} describes s-wave superconductivity. 
We consider an isolated vortex in the center of a two-dimensional disc with radius $R$ and assume 
$\Delta(\bm{r}) = \Delta(r) e^{\ii \kappa \theta}$, where $\kappa$ is the vorticity 
and takes the value $\pm1$. We also assume that $\Delta(r)$ approaches a constant value $\Delta_{0}(>0)$ far from the vortex core. When $V_{z}^{2}>\mu^{2}+\Delta_{0}^{2}$ is satisfied, the system is in the topological phase and Majorana zero-energy states appear in the vortex core and near the edge.
From the above Hamiltonian, we can obtain the following BdG equation:
\begin{align}
&\check{H}_{\BdG} (\bm{r})\vec{u}_{K}(\bm{r}) = E_{K}\vec{u}_{K}(\bm{r}), \\
&\check{H}_{\BdG}(\bm{r}) = \begin{bmatrix}\hat{H}_{0} & \hat{\Delta}(\bm{r}) \\ \hat{\Delta}^{\dagger}(\bm{r}) & -\hat{H}_{0}^{*} \end{bmatrix}
, \  \hat{\Delta}(\bm{r}) = \Delta(\bm{r})\ii \hat{\sigma}_{y}. \label{Hbdg}
\end{align}
Here, $\vec{u}_{K}=\trans(u_{K\uparrow},u_{K\downarrow},v_{K\uparrow},v_{K\downarrow})$ is a four-component vector. 
The angular momenta are good quantum numbers since the system has rotational symmetry. 
The eigenvector can be decomposed into angular and radial parts as $\vec{u}_{K=(l,\nu)}(\bm{r}) = \check{U}_{l}(\theta)\vec{u}_{l,\nu}(r)$ with 
$\check{U}_{l}(\theta) = \mathrm{diag}(e^{\ii l\theta},e^{\ii (l+1)\theta},e^{\ii l\theta},e^{\ii (l-1)\theta})/\sqrt{2\pi}$ for $\kappa = 1$ 
and $\check{U}_{l}(\theta) = \mathrm{diag}(e^{\ii (l-1)\theta},e^{\ii l\theta},e^{\ii (l+1)\theta},e^{\ii l\theta})/\sqrt{2\pi}$ for $\kappa = -1$. 
Here, we take $\nu$ as the radial quantum number. 
We can expand this radial part of the eigenstate $\vec{u}_{l,\nu}(r)$ by the Fourier-Bessel expansion~\cite{PhysRevB.82.174506} 
and diagonalize the matrix for each angular momentum $l$ to obtain the sets of eigenvalues and eigenvectors. 
(Subscripts of the matrix represent the indices of the zeroth points of Bessel functions.) 
We find that there are two low-energy modes with energies below the gap in the bulk $\Delta_{\bb}$; 
one is localized in the vortex core and the other is near the edge. We label them as $\nu = \mathrm{c}$ and $\nu = \mathrm{e}$ respectively. 
(The definition of $\Delta_{\bb}$ is mentioned in Sect.~\ref{sec-num}.)
We remark on the two zero-energy states. 
In this numerical calculation, we consider finite-size but sufficiently large systems; 
hence, the two zero-energy states have finite but small energies and their signs are opposite. 
We specify them by $\nu = +$ and $\nu = -$. 
Each wave function has localized distributions around the vortex core and the edge simultaneously.
We can divide them by taking a suitable linear combination of these two states 
and call the state bound in the vortex core $\nu = \mathrm{c}$.

\begin{figure}
\begin{center}
\includegraphics[width=7cm]{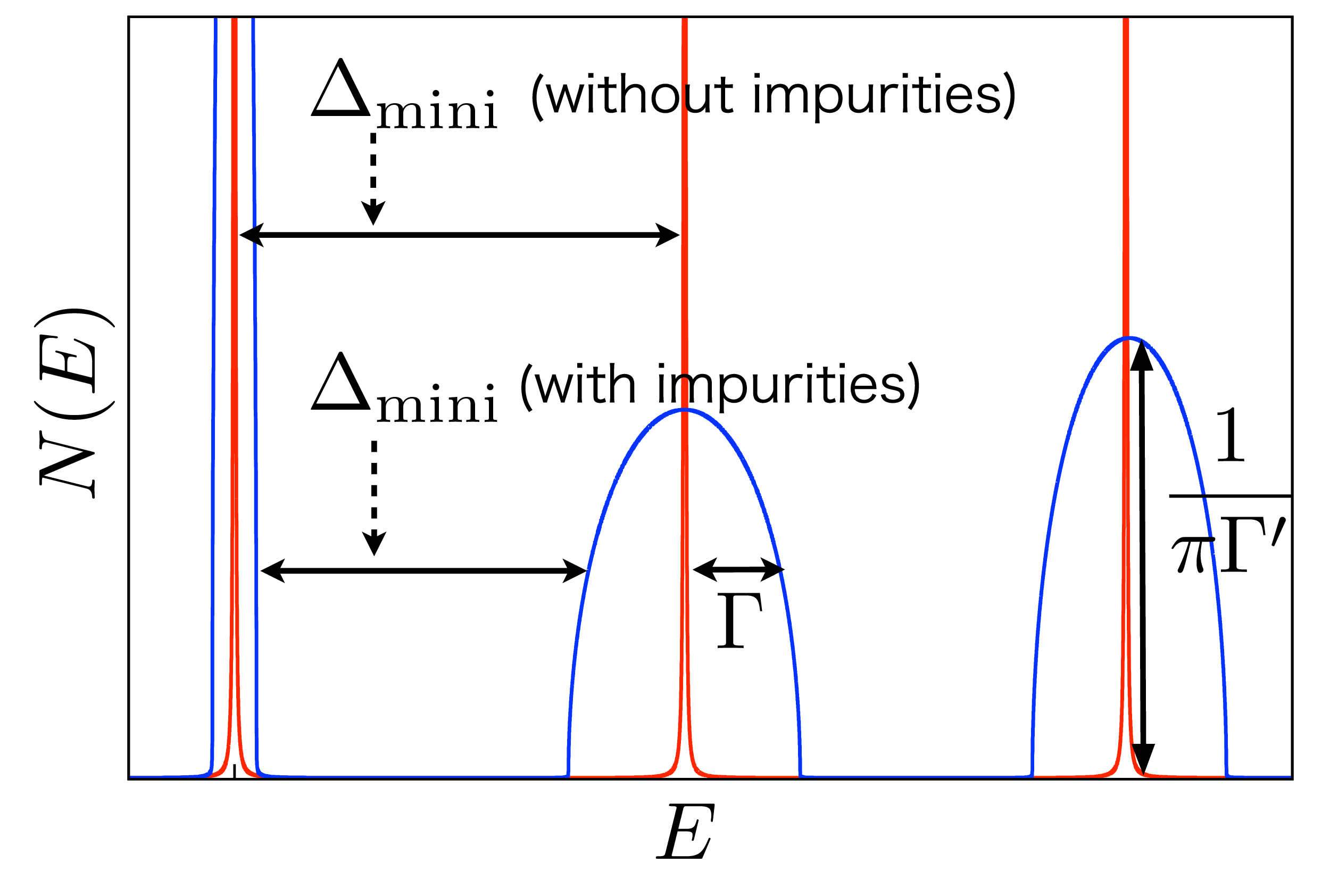}
\caption{(Color online) Relationships between spectra, impurity scattering rates, and minigap are shown schematically.}
\label{scheme}
\end{center}
\end{figure}
We calculate the scattering rates of the vortex core states within the single-mode approximation using only the mode $\nu = \mathrm{c}$ obtained above.
We consider that there exist weak but many scatterers and treat them within the SCBA. 
The scheme used to calculate the scattering rates is given in the following form as discussed in Ref.~\citen{masakikato}:
\begin{align}
\check{G}(\bm{r},\bm{r}';\ii\omega_{n}) = \sum_{l}\dfrac{\check{\tau}_{z}\vec{u}_{l,\mathrm{c}}(\bm{r})(\vec{u}_{l,\mathrm{c}}(\bm{r}'))^{\dagger}}{E_{l,\mathrm{c}}-\sigma_{l}(\ii\omega_{n})-\ii\omega_{n}}, \label{green}\\
\sigma_{l}(\ii\omega_{n}) = \dfrac{\Gamma_{\mathrm{n}}}{2\pi^{2}N_{\mathrm{n}}}\sum_{l'}\dfrac{M_{l,l'}}{E_{l',\mathrm{c}}-\sigma_{l'}(\ii\omega_{n})-\ii \omega_{n}}, \label{sigma}\\
M_{l,l'} = \int r dr \left|\left(\vec{u}_{l,\mathrm{c}}(r)\right)^{\dagger}\check{\tau}_{z}\vec{u}_{l',\mathrm{c}}(r)\right|^{2}, \label{mat-el}
\end{align}
where $\check{\tau}_{z}=\mathrm{diag}(1,1,-1,-1)$ and $\vec{u}_{l,\mathrm{c}}(r)$ denotes the radial part of $\vec{u}_{l,\mathrm{c}}(\bm{r})$.
$\Gamma_{\mathrm{n}}$ and $N_{\mathrm{n}}$ are, respectively, the impurity scattering rates and the DoS per spin at the Fermi energy $\epsilon_{\F}$ in the normal state.
$\omega_{n}$ is the fermion Matsubara frequency, and in the following, we perform the analytical continuation $\ii\omega_{n} \to \omega +\ii \delta$.
The DoS can be obtained from Green's function:
\begin{align}
N(\omega) = \sum_{l} N_{l}(\omega) = \int d\bm{r}\mathrm{Im}\left[\dfrac{1}{\pi}\mathrm{Tr}\left.\check{\tau}_{z}\check{G}(\bm{r},\bm{r};\ii\omega_{n})\right|_{\ii\omega_{n}\to\omega+\ii\delta}\right].
\label{dos}
\end{align}
We define the impurity scattering rates (denoted as $\Gamma$) in two ways. 
The first definition of $\Gamma$ is the half width at half maximum of the spectrum, 
shown as $\Gamma$ in Fig.~\ref{scheme}, and we use this definition in Sect.~\ref{sec-num}.
The second definition of $\Gamma$ is 
the inverse of the DoS multiplied by $\pi$ $[\Gamma=(\mathrm{Im}G)^{-1}]$, 
shown as $\Gamma'$ in Fig.~\ref{scheme}, and we use this definition in Sect.~\ref{sec-andreev}.
These definitions are equivalent when the spectrum takes the Lorentzian profile.

\begin{figure}
\begin{center}
\includegraphics[width=7cm]{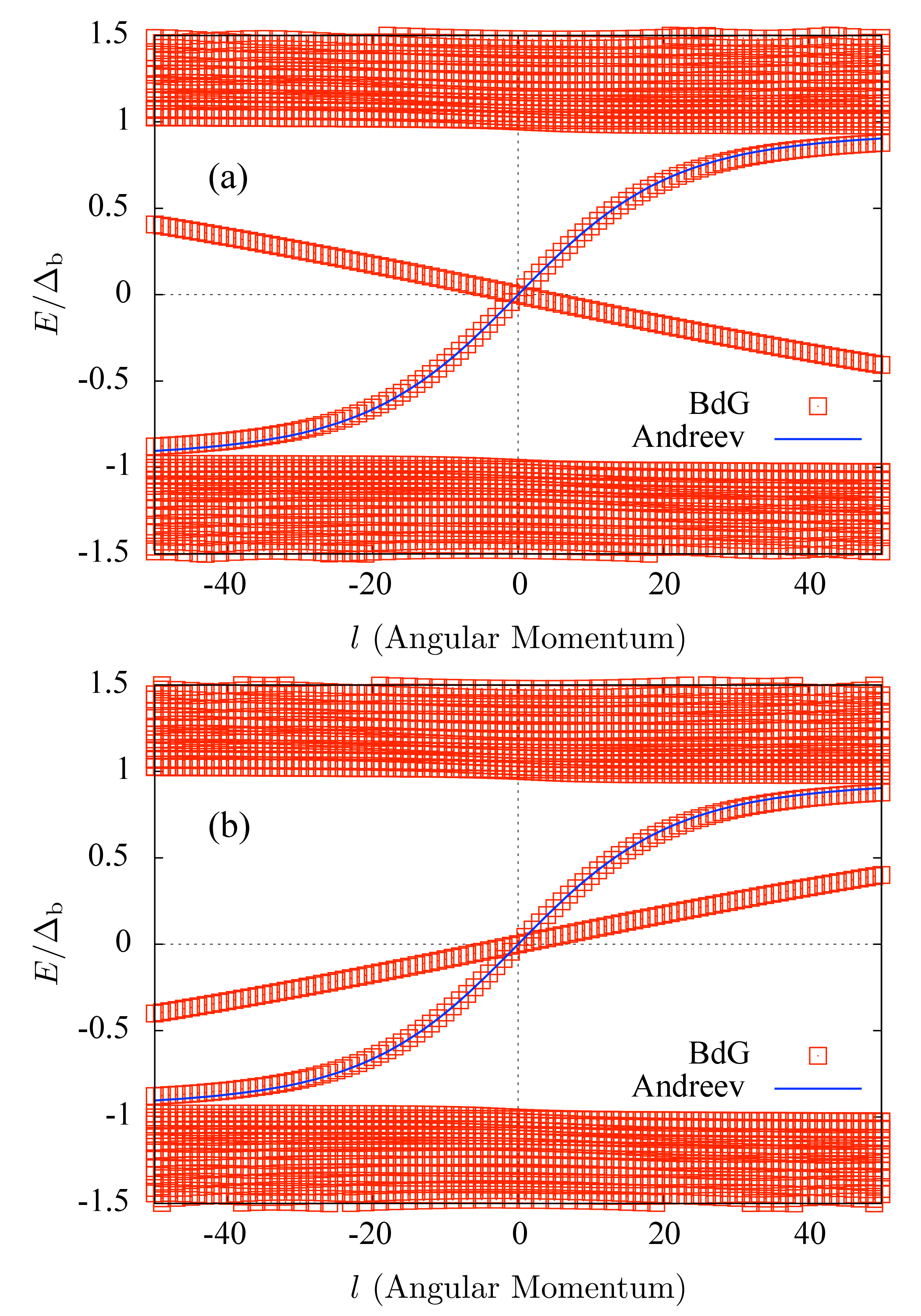}
\caption{(Color online) Energy spectrum for $\mu=0$, $k_{\F}\xi = 20$, and $m\alpha^{2}/|V_{z}|=1.6$. 
(a) The sign of the Zeeman coupling is positive and the slopes of the two low-energy branches are opposite in sign. 
(b) The sign of the Zeeman coupling is negative and the slopes have the same sign. The blue solid line labeled as ``Andreev'' is the analytical solution for the CdGM mode discussed in Sect.~\ref{sec-andreev}.}
\label{fig-E}
\end{center}
\end{figure}

\section{Numerical Calculation}\label{sec-num}
In this section, we show the results of numerical calculations for scattering rates. 
First, we explain the parameters used in our calculations.
We fix the quasiclassical parameter $k_{\F}\xi=20$, where $k_{\F}$ and $\xi$ are the Fermi momentum and  coherence length, respectively.
In order to set this parameter, we consider the homogeneous system with the s-wave pair potential $\Delta(\bm{r}) = \Delta_{0}$ for the moment.
In the topological phase, the Fermi momentum $k_{\F}$ and the Fermi velocity $v_{\F}$ are well defined 
since this two-band model has a single Fermi surface for the uniform system. 
We set $\Delta_{\bb}$ as the minimum gap of the one-particle excitation spectrum around $k=k_{\F}$ and 
define the coherence length as $\xi  \equiv v_{\F}/\Delta_{\bb} $ with this minimum excitation gap $\Delta_{\bb}$. 
[Note the difference between $\Delta_{0}$ and $\Delta_{\bb}$; 
$\Delta_{0}$ is simply the amplitude of the s-wave pair potential and $\Delta_{\bb}$ is the minimum excitation gap. 
In general, they are not the same and 
approximately satisfy $|\Delta_{\bb}/\Delta_{0}| \simeq |\alpha| k_{\F}/(V_{z}^{2}+\alpha^{2}k_{\F}^{2})^{1/2}$ as Eq.~\eqref{eq-mini}.]
We set $\mu = 0$ and $\Delta_{0}$ as the unit of energy and change the ratio of $m\alpha^{2}/|V_{z}|$ under the condition 
$V_{z}^{2}>\mu^{2}+\Delta_{0}^{2}$. 
We remark again that this system has two inequivalent vortices in terms of the symmetry operation. 
We can distinguish this inequivalence by 
the relative sign of the slopes of the vortex core mode and chiral edge mode: 
$\partial E_{l,\mathrm{c}}/\partial l$ and $\partial E _{l,\mathrm{e}}/\partial l$, as shown in Fig.~\ref{fig-E}. 
We also find that the signs of the Zeeman coupling $V_{z}$ and the vorticity correspond to the signs of the slopes of the chiral edge mode and vortex core mode, respectively~\cite{Wu2012}. 
This structure is very similar to that of the chiral p-wave SC if we regard the sign of the slope of the edge mode as the chirality of the Cooper pair. 
We thus call the type of vortex with the positive (negative) relative sign the ``parallel (antiparallel) vortex''.
In this system, the signs of the two slopes are the same (opposite) when $\kappa V_{z}>0 \ (\kappa V_{z}<0)$,  
and in the following numerical calculation, we only change the sign of the Zeeman coupling and 
fix the vorticity as $\kappa = -1$.
To solve the BdG equation, we set the system size to $R= 20\xi$ and the spatial profile of the pair potential in the radial direction to $\Delta(r) = \Delta_{0}\tanh(r/\xi)$. 
Moreover, we introduce two cutoffs, $l_{c} = 50$ and $N_{c}=400$, which are the numbers of angular momentum as a quantum number and zero points of the Bessel function, respectively, and the infinitesimal quantity $\delta = 10^{-6}\Delta_{\bb}$ in Eq.~\eqref{dos}. 
Under these parameters, we numerically solve the BdG equation and Eqs.~\eqref{green}~--~\eqref{dos} 
to calculate the DoS and scattering rates in the presence of impurities.

\begin{figure}
\begin{center}
\includegraphics[width=7cm]{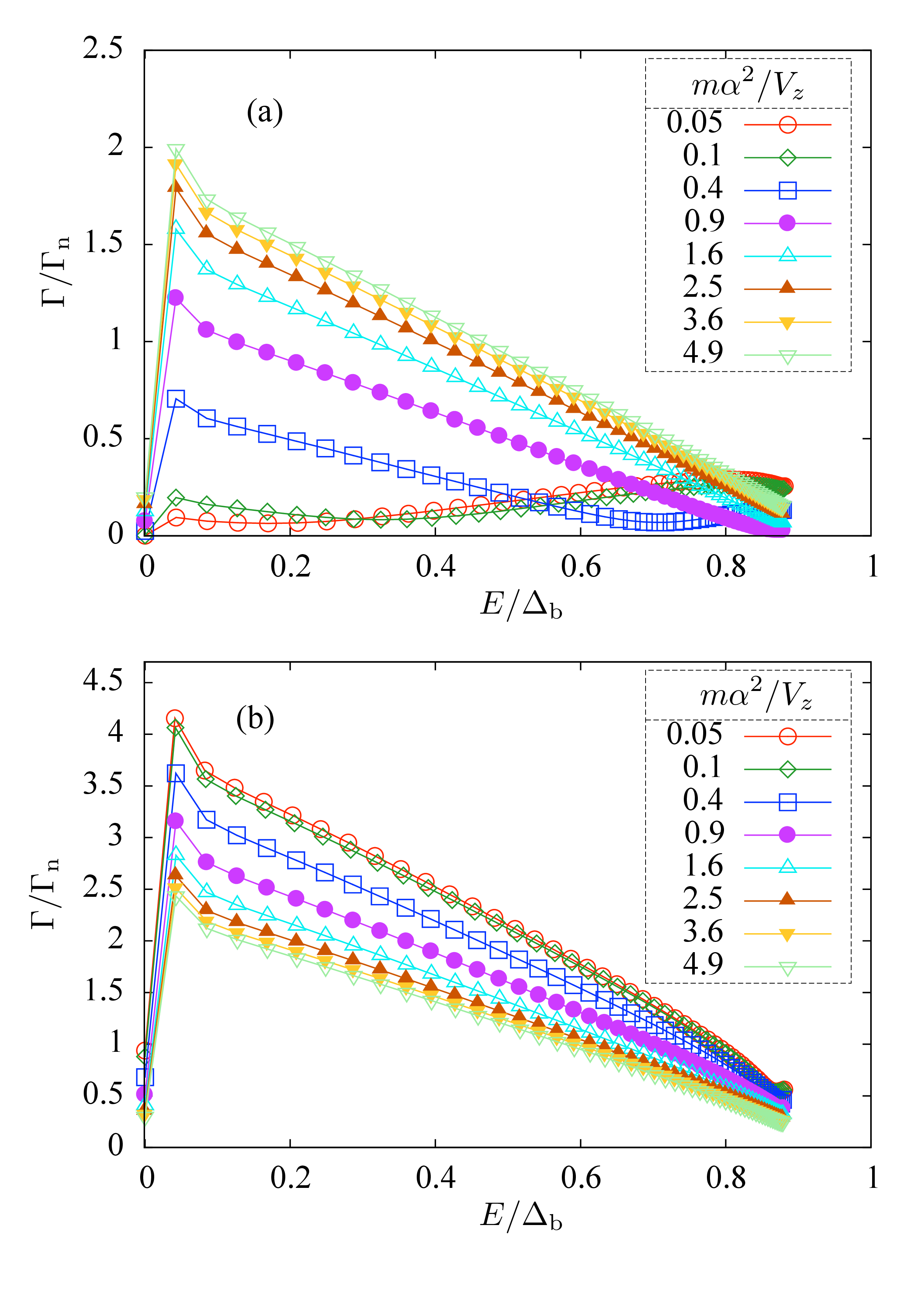}
\caption{(Color online) Impurity scattering rates for various values of $m\alpha^{2}/V_{z}$ and $\Gamma_{\n}/\pi \Delta_{\bb} = 10^{-3}$. (a) $V_{z}>0$ (antiparallel vortex) and (b) $V_{z}<0$ (parallel vortex).}
\label{fig-gamma}
\end{center}
\end{figure}

We show the scattering rates $\Gamma$ for various values of $m\alpha^{2}/|V_{z}|$ 
in Figs.~\ref{fig-gamma}(a) and \ref{fig-gamma}(b), which correspond to the positive and negative Zeeman couplings, respectively. 
The horizontal axis shows the eigenenergy for pure systems, which is scaled by the gap $\Delta_{\mathrm{b}}$ for each parameter. 
The discrete eigenenergies scaled by $\Delta_{\bb}$ are independent of the ratio $m\alpha^{2}/V_{z}$, 
and the level spacings scaled by $\Delta_{\bb}$ in the low-energy region are the inverse of $k_{\F}\xi$.
We set $\Gamma_{\mathrm{n}} = 10^{-3} \pi \Delta_{\mathrm{b}}$. 
Note again that $\Delta_{\mathrm{b}}$ is determined by some parameters such as $V_{z}$, $m\alpha^{2}$, and so on.

These figures show that the scattering rates are smaller for $V_{z}>0$ than for $V_{z}<0$. 
One might consider that the impurity effects are characterized by the type of vortex (parallel or antiparallel), 
because in the chiral p-wave SC, the bound states in an antiparallel vortex are robust against impurities, 
while those in a parallel vortex are sensitive to impurities. 
It is impossible to characterize the impurity effects only by means of the relative sign of the vorticity and chirality since there are finite scattering rates even for the antiparallel vortex, as shown in Fig.~\ref{fig-gamma}(a).
We need to consider the dependence of scattering rates on another parameter $m\alpha^{2}/V_{z}$.
We find that the scattering rates in the low-energy region are a decreasing function of the ratio 
$m\alpha^{2}/|V_{z}|$ for the parallel vortex [Fig.~\ref{fig-gamma}(b)], 
while they are increasing function for the antiparallel vortex [Fig.~\ref{fig-gamma}(a)].
In particular, for very small $m\alpha^{2}/|V_{z}|$ of $0.05$ and $0.1$,
the scattering rates are exceptionally suppressed for the antiparallel vortex, and the difference between the two inequivalent vortices is evident, similarly to the chiral p-wave SC.
On the other hand, when $m\alpha^{2}/|V_{z}|$ is large, 
the scattering rates of the two vortices have almost the same energy dependence and magnitude, 
which is a different feature from the case of the chiral p-wave SC.
(We comment on the exceptionally small scattering rates of zero-energy states in spite of the finite scattering rates of other excited states. This is independent of the type of vortex or the parameter $m\alpha^{2}/V_{z}$.)

\begin{figure}
\includegraphics[width=7cm]{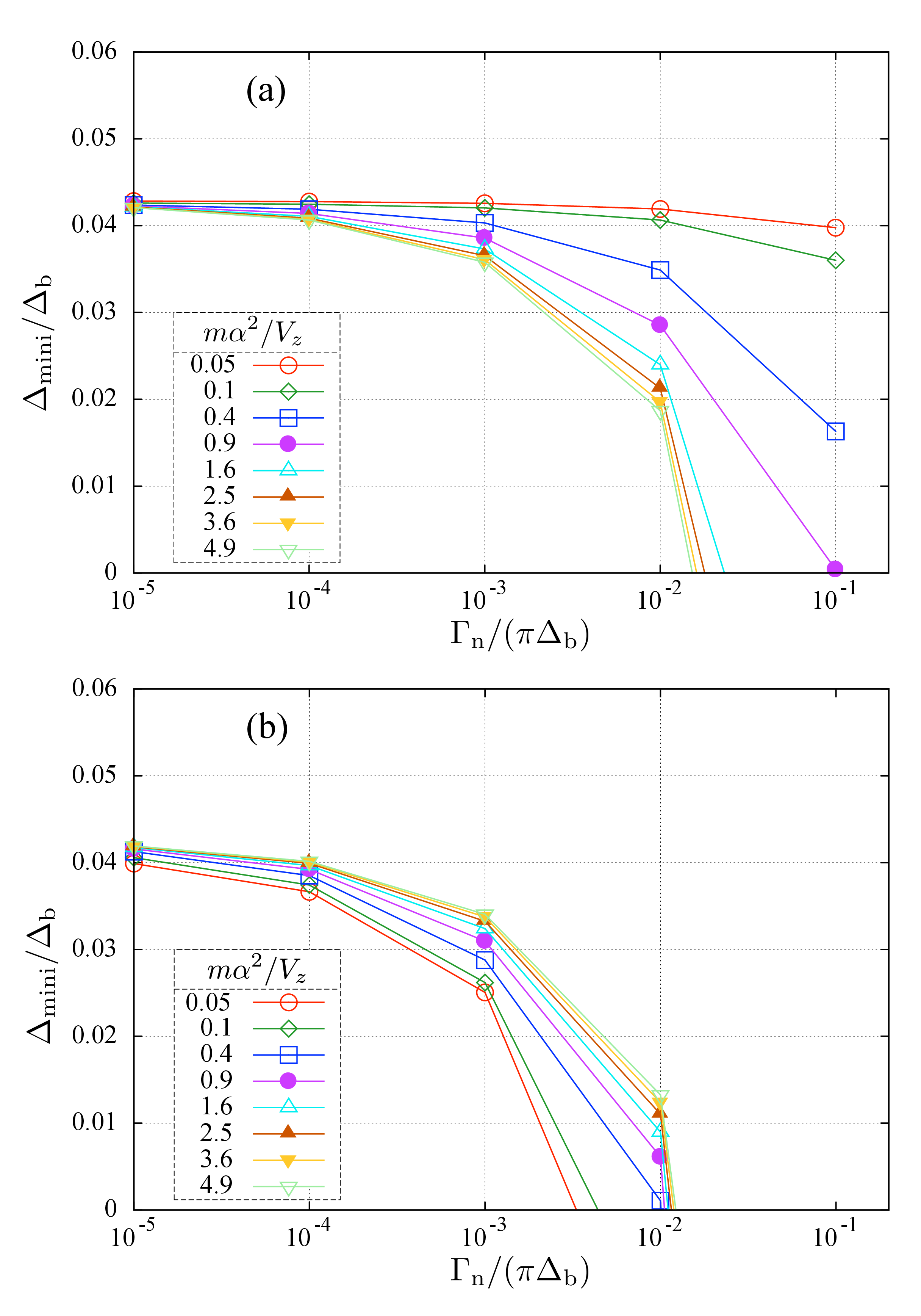}
\caption{(Color online) Impurity effects on minigap. The horizontal axis denotes the impurity strength $\Gamma_{\n}/(\pi\Delta_{\bb})$. The symbols have the same meanings as those in Fig.~\ref{fig-gamma}. }
\label{fig-minigap}
\end{figure}

We note again that the adiabaticity of TQC requires the robustness of minigaps, and thus we show the $\Gamma_{\n}$ dependence of the minigap 
in Fig.~\ref{fig-minigap}(a) for $V_{z}>0$ and in Fig.~\ref{fig-minigap}(b) for $V_{z}<0$.  
We define the minigap in the presence of impurities by subtracting the widths of the first excited state and zero mode from the minigap for a pure system, as shown in Fig.~\ref{scheme}. 
Impurity effects on minigaps reflect the scattering rates in low-energy states. 
In Fig.~\ref{fig-minigap}(a), only for the case of small $m\alpha^{2}/|V_{z}|$ with $V_{z}>0$, 
the minigap is robust against the impurities; 
otherwise, the size of the minigap steeply decreases with increasing $\Gamma_{\n}$. 
In this system, 
the stability of the minigap crucially depends on the type of vortices and the parameter $m\alpha^{2}/V_{z}$.

We consider the origin of the dependence of the scattering rates on the type of vortex and the parameter 
within the Andreev approximation in the next section.

\section{Andreev Approximation}\label{sec-andreev}
\subsection{Spectrum and wave functions of low-energy states}
In this section, we analytically study the bound states in the vortex core within the Andreev approximation for the BdG equation. 
Tewari et al. also discussed the zero-energy states in the vortex core by solving the approximated BdG equation analytically~\cite{Tewari2010219}.
The main strategy of the approximation is the same as in the following discussion, 
but the details of the calculation are slightly different.
Moreover, our aim is not only the construction of a zero-energy state but also the construction of nonzero-energy states and the understanding of the scattering rates.
In Appendix\ref{ap-edge}, we also discuss the possibility of the construction of the edge state within the Andreev approximation.

We start with the BdG equation for the uniform case.
\begin{align}
\check{H}_{\mathrm{BdG}}(\bm{k}) \vec{\bar{u}}_{\bm{k}} = E \vec{\bar{u}}_{\bm{k}}, \label{bdg_uni}
\end{align}
where $\check{H}_{\mathrm{BdG}}(\bm{k})$ is a 4 by 4 matrix and described as 
\begin{align}
\check{H}_{\mathrm{BdG}} (\bm{k})=\begin{bmatrix}
\hat{H}_{0}(\bm{k})  & \hat{\Delta} \\
\hat{\Delta}^{\dagger} & -\hat{H}_{0}^{*}(-\bm{k})
\end{bmatrix}, \\
\hat{H}_{0}(\bm{k})=\begin{bmatrix}
\epsilon_k +V_z& \ii \alpha k e^{-\ii \phi_{\bm{k}}}\\
-\ii \alpha ke^{\ii \phi_{\bm{k}}}&\epsilon_k-V_z\end{bmatrix}, \ 
\hat{\Delta} = \begin{bmatrix} 0 & \Delta \\ -\Delta & 0\end{bmatrix}, \\
\epsilon_{k} = \dfrac{k^{2}}{2m}-\mu, \quad \bm{k} = k\cos(\phi_{\bm{k}})\bm{e}_{x} + k\sin(\phi_{\bm{k}})\bm{e}_{y}.
\end{align}
$\vec{\bar{u}}_{\bm{k}}$ is a four-component vector and taken as $\vec{u}_{\bm{k}} \equiv e^{\ii \bm{k}\cdot \bm{r}} \vec{\bar{u}}_{\bm{k}}$. $\vec{u}_{\bm{k}}$ is a spatially oscillating wave function.

First of all, we take the unitary transformation $\vec{u}_{\bm{k}} = \check{U}(\bm{k}) \vec{u}_{\bm{k}}^{\bb} 
= e^{\ii \bm{k}\cdot \bm{r}} \check{U}(\bm{k}) \vec{\bar{u}}_{\bm{k}}^{\bb} $, in which basis $\hat{H}_{0}(\bm{k})$ is a diagonal matrix.
One of the unitary matrices is taken as
\begin{align}
\check{U}(\bm{k}) &= \begin{bmatrix}\hat{U}_{1}(\bm{k}) & \hat{0}\\ \hat{0} & \hat{U}_{1}^{*}(-\bm{k})\end{bmatrix}, \\
\hat{U}_{1}(\bm{k}) &= \dfrac{1}{C}\begin{bmatrix}\ii \alpha k e^{-\ii \phi_{\bm{k}}} & \sqrt{V_{z}^{2}+\alpha^{2}k^{2}}-V_{z}\\
\sqrt{V_{z}^{2}+\alpha^{2}k^{2}}-V_{z}& \ii \alpha k e^{\ii \phi_{\bm{k}}}
\end{bmatrix}, \\
C^{2} &= 2 \sqrt{V_{z}^{2}+\alpha^{2}k_{\F}^{2}}\left(\sqrt{V_{z}^{2}+\alpha^{2}k_{\F}^{2}}-V_{z}\right). \label{u-norm}
\end{align}
Through this transformation, we obtain the following Hamiltonian~\cite{PhysRevB.81.125318}:
\begin{align}
\check{H}_{\BdG}^{\bb}(\bm{k}) \vec{\bar{u}}_{\bm{k}}^{\bb} = E\vec{\bar{u}}_{\bm{k}}^{\bb}, \label{band_bdg}\\
\check{H}_{\BdG}^{\bb}(\bm{k}) = \check{U}^{\dagger}(\bm{k})\check{H}_{\BdG}(\bm{k})\check{U}(\bm{k}) =
\begin{bmatrix}
\hat{H}_{0}^{\bb} &\hat{\Delta}^{\bb} \\
\left(\hat{\Delta}^{\bb}\right)^{\dagger} & -\hat{H}_{0}^{\bb} \\
\end{bmatrix}, \\
\hat{H}_{0}^{\bb} = \mathrm{diag}(E_{k+},E_{k-}), E_{k\pm} = \epsilon_{k} \pm \sqrt{V_{z}^{2}+\alpha^{2}k^{2}},\\ \hat{\Delta}^{\bb} = 
\begin{bmatrix}
-\ii f_{p}(k)\Delta e^{\ii\phi_{\bm{k}}} & f_{s}(k)\Delta \\
-f_{s}(k)\Delta& \ii f_{p}(k)\Delta e^{-\ii\phi_{\bm{k}}}
\end{bmatrix}, \\
f_{s}(k) = \dfrac{V_{z}}{\sqrt{V_{z}^{2}+\alpha^{2}k^{2}}},\ f_{p}(k) = \dfrac{\alpha k}{\sqrt{V_{z}^{2}+\alpha^{2}k^{2}}}.
\end{align}
For $V_{z}^{2}>\Delta^{2}+\mu^{2}$, we find $E_{k+}>0$ and the existence of $k_{\F}$ such that $E_{k_{\F}-} = 0$. 
We expand the above BdG equation [Eq.~\eqref{band_bdg}] around $k_{\F}$ 
while retaining the leading-order terms only, 
i.e., up to the first order for $E_{k-}$ and the zeroth order for the other elements. 
We call this approximation the Andreev approximation. We introduce two quantities as follows:
\begin{align}
E_{k+} &\simeq  E_{k_{\F}+} = 2\sqrt{V_{z}^{2}+\alpha^{2}k_{\F}^{2}}\equiv E_{c}, \label{E+}\\
E_{k-}&\simeq \dfrac{\partial E_{k-}}{\partial \bm{k}}\cdot(\bm{k}-\bm{k_{\F}}) \nonumber \\
&= \left(1-\dfrac{m\alpha^{2}}{\sqrt{V_{z}^{2}+\alpha^{2}k_{\F}^{2}}}\right)\dfrac{\bm{k}_{\F}}{m}  \cdot(\bm{k}-\bm{k_{\F}}) \equiv \bm{v}_{\F}\cdot(\bm{k}-\bm{k_{\F}}). \label{E-}
\end{align}
Here, the Fermi velocity $\bm{v}_{\F}$ is parallel to the Fermi momentum $\bm{k}_{\F}$. 
In the following part, we consider a single vortex in the system and set $\Delta = \Delta(r)e^{\ii \kappa\theta}$ 
($\kappa$: vorticity). 
We define a 2 by 2 matrix $R(\theta)$ as the rotation around the $z$ axis by angle $\theta$ 
and introduce the basis of the 2D polar coordinates as $(\bm{e}_{r},\bm{e}_{\theta}) = R(\theta)(\bm{e}_{x},\bm{e}_{y})$. 
In the bulk region far from the vortex core, $\Delta(r)$ approaches $\Delta_{0} (>0)$, which is the magnitude of the s-wave pair potential. 
Because of this inhomogeneity, we replace $(\bm{k}-\bm{k}_{\F})$ by $-\ii \bm{\nabla}$ in Eq.~\eqref{E-}. 
In the following, we simply use $f_{p}$ and $f_{s}$ instead of $f_{p}(k_{\F})$ and $f_{s}(k_{\F})$, respectively.
We can thus write down the differential equation for the slowly varying function $\vec{\bar{u}}_{\bm{k}}^{\bb}$ as 
\begin{align}
E_{c} \bar{u}^{\bb}_{\bm{k}_{\F}+} -\ii f_{p}e^{\ii \phi_{\bm{k}_{\F}}} \Delta\bar{v}^{\bb}_{\bm{k}_{\F}+} 
+f_{s}\Delta \bar{v}^{\bb}_{\bm{k}_{\F}-} =E \bar{u}^{\bb}_{\bm{k}_{\F}+}, \\
-\ii \bm{v}_{\F}\cdot \nabla \bar{u}^{\bb}_{\bm{k}_{\F}-}  
- f_{s}\Delta \bar{v}^{\bb}_{\bm{k}_{\F}+} 
+\ii f_{p}e^{-\ii \phi_{\bm{k}_{\F}}} \Delta\bar{v}^{\bb}_{\bm{k}_{\F}-}  =E \bar{u}^{\bb}_{\bm{k}_{\F}-},\\
-E_{c} \bar{v}^{\bb}_{\bm{k}_{\F}+}  +\ii f_{p}e^{-\ii \phi_{\bm{k}_{\F}}} \Delta^{*}\bar{u}^{\bb}_{\bm{k}_{\F}+} 
-f_{s}\Delta^{*} \bar{u}^{\bb}_{\bm{k}_{\F}-} =E \bar{v}^{\bb}_{\bm{k}_{\F}+}, \\
\ii \bm{v}_{\F}\cdot \nabla  \bar{v}^{\bb}_{\bm{k}_{\F}-}  
+f_{s}\Delta^{*} \bar{u}^{\bb}_{\bm{k}_{\F}+}
-\ii f_{p}e^{\ii \phi_{\bm{k}_{\F}}} \Delta^{*}\bar{u}^{\bb}_{\bm{k}_{\F}-} 
 =E \bar{v}^{\bb}_{\bm{k}_{\F}-}. 
\end{align}
We can easily eliminate $\bar{u}^{\bb}_{\bm{k}_{\F}+}$ and $\bar{v}^{\bb}_{\bm{k}_{\F}+}$ and obtain the following two coupled differential equations:
\begin{align}
&\left[-\ii \bm{v}_{\F}\cdot \bm{\nabla}\hat{\sigma}_{z}-\mathrm{Re}\Delta_{p2}\hat{\sigma}_{x}
+\mathrm{Im}\Delta_{p2}\hat{\sigma}_{y}\right.\nonumber \\ \
&\hspace{8em}\left.
+\hat{\Delta}_{s}\hat{A}\hat{\Delta}_{s}\right]\begin{pmatrix}
\bar{u}^{\bb}_{\bm{k}_{\F}-} \\ \bar{v}^{\bb}_{\bm{k}_{\F}-}
\end{pmatrix}
= E \hat{1} 
\begin{pmatrix}
\bar{u}^{\bb}_{\bm{k}_{\F}-} \\ \bar{v}^{\bb}_{\bm{k}_{\F}-}
\end{pmatrix}, \label{Andreev1}
\end{align}
where we define $\Delta_{p1,2} = -\ii f_{p}e^{\pm\ii\phi_{\bm{k}_{\F}}}\Delta,\ \Delta_{s} = f_{s}\Delta$ and 
\begin{align}
\hat{A} &=\dfrac{1}{E^{2}-E_{c}^{2}-|\Delta_{p1}|^{2}}\begin{bmatrix}E + E_{c} & \Delta_{p1} \\ \Delta_{p1}^{*} & E- E_{c} \end{bmatrix}, \nonumber \\ 
\hat{\Delta}_{s}&=-\ii\begin{bmatrix}
0 & \Delta_{s}\\
-\Delta_{s}^{*} & 0 
\end{bmatrix}. \nonumber 
\end{align}
We can construct the zero-energy state ($E = 0$) analogous to the Jackiw-Rebbi solution for the massive Dirac equation~\cite{PhysRevD.13.3398}
and obtain the low-energy spectrum through the first-order perturbation even in the presence of $\hat{A}$. 
However, the contribution from $\hat{A}$, i.e., $(\bar{u}_{\bm{k}_{\F}+}^{\bb},\bar{v}_{\bm{k}_{\F}+}^{\bb})$, is small compared with that from $(\bar{u}_{\bm{k}_{\F}-}^{\bb},\bar{v}_{\bm{k}_{\F}-}^{\bb})$ 
and we can thus omit the fourth term on the left-hand side. of Eq.~\eqref{Andreev1} and hereafter analyze the following equation:
\begin{align}
\left[-\ii \bm{v}_{\F}\cdot \bm{\nabla}\hat{\sigma}_{z}-\mathrm{Re}\Delta_{p2}\hat{\sigma}_{x}
+\mathrm{Im}\Delta_{p2}\hat{\sigma}_{y}
\right]\begin{pmatrix}
\bar{u}^{\bb}_{\bm{k}_{\F}-} \\ \bar{v}^{\bb}_{\bm{k}_{\F}-}
\end{pmatrix}
= E \hat{1} 
\begin{pmatrix}
\bar{u}^{\bb}_{\bm{k}_{\F}-} \\ \bar{v}^{\bb}_{\bm{k}_{\F}-}
\end{pmatrix}. \label{Andreev2}
\end{align}
We introduce another set of coordinates $(s,b)$ as $(\bm{e}_{s},\bm{e}_{b}) = R(\phi_{\bm{k}})(\bm{e}_{x},\bm{e}_{y})$.
In this frame, $\bm{v}_{\F} = v_{\F}\bm{e}_{s}$. We take the gauge transformation as 
$\trans(\bar{u}_{\bm{k}_{\F}-}^{\bb},\bar{v}_{\bm{k}_{\F}-}^{\bb}) 
= \exp[\ii (\kappa-1)\phi_{\bm{k}_{\F}}\hat{\sigma}_{z}/2]
\trans(\tilde{u}_{\bm{k}_{\F}-}^{\bb},\tilde{v}_{\bm{k}_{\F}-}^{\bb})$; 
then Eq.~\eqref{Andreev2} is reduced to
\begin{align}
\left[-\ii v_{\F} \partial_{s} \hat{\sigma}_{z}-f_{p}\Delta(r)\dfrac{s}{r}\hat{\sigma}_{y}-\kappa f_{p}\Delta(r)\dfrac{b}{r}\hat{\sigma}_{x}\right]
\begin{pmatrix}\tilde{u}_{\bm{k}_{\F}-}^{\bb}\\ \tilde{v}_{\bm{k}_{\F}-}^{\bb}\end{pmatrix}= E\hat{1}\begin{pmatrix}\tilde{u}_{\bm{k}_{\F}-}^{\bb}\\ \tilde{v}_{\bm{k}_{\F}-}^{\bb}\end{pmatrix} \label{Andreev3}.
\end{align}
Note that $r = \sqrt{s^{2}+b^{2}}$ and $\tan(\theta-\phi_{k_{\F}}) = b/s$. 
$b$ is the impact parameter in scattering theory and related to the angular momentum $l$ through $l = -k_{\F}b$.
From this point, 
we find that there exists a zero-energy state when $b=0$.
For small $b$, we treat  the third term of Eq.~\eqref{Andreev3} as a perturbation Hamiltonian and obtain other low-energy states. 
We can construct the solution with the energy $E=0$ for the unperturbed Hamiltonian:
\begin{align}
\begin{pmatrix}\tilde{u}_{\bm{k}_{\F}-}^{\bb}\\ \tilde{v}_{\bm{k}_{\F}-}^{\bb}\end{pmatrix} = \exp\left[-\psi(s,b)\right]\begin{pmatrix}1 \\ -\mathrm{sgn}(\alpha)\end{pmatrix},\\
\psi(s,b) =\dfrac{1}{v_{\F}}\int_{0}^{s} ds' \left(\left|f_{p}\right|\Delta(r')\dfrac{s'}{r'}\right),
\end{align}
where $r' = \sqrt{s'^{2}+b^{2}}$. 
With this wave function, we can obtain the correction of energy through the first-order perturbation:
\begin{align}
E(b) &= \Delta E_{1} =\dint_{0}^{\infty}ds \kappa \left|f_{p}\right|\Delta(r)(b/r)\exp\left[-2\psi(s,b)\right]/N\xi, \label{energy-spec}\\
N &\equiv \dint_{0}^{\infty}ds\exp\left[-2\psi(s,b)\right]/\xi. \label{Norm}
\end{align}
The solid lines shown in Fig.~\ref{fig-E} represent the energy spectrum given by Eq.~\eqref{energy-spec} for $k_{\F}\xi=20$ and $m\alpha^{2}/|V_{z}| = 1.6$. 
For this parameter, Eq.~\eqref{energy-spec} is in good agreement with the energy spectrum numerically obtained from the BdG equation. 
For a smaller $m\alpha^{2}/|V_{z}|$, the deviation of the energy spectrum for a large angular momentum is larger, but the low-energy spectrum can be described well.
From this, an expression for the minigap can be approximately obtained:
\begin{align}
\Delta_{\mathrm{mini}} \equiv \left|\left.\dfrac{1}{k_{\F}}\dfrac{\partial E}{\partial b}\right|_{b=0}\right|= c_{1}\dfrac{\left|f_{p}\right|\Delta_{0}}{k_{\F}\xi}
\simeq c_{1}\dfrac{\Delta_{\bb}}{k_{\F}\xi}, \label{eq-mini}
\end{align}
where $c_{1}$ is a constant on the order of unity and, in particular, $c_{1} = 7\zeta(3)/\pi^{2}$ when $\Delta(r) =\Delta_{0} \tanh (r/\xi)$.

\subsection{Scattering rates}
Hereafter, making use of the above solutions to the Andreev equation,
we calculate the scattering rates for low-energy states. 
In the following calculation, as mentioned in Sect.~\ref{sec-model-method}, we change the definition of $\Gamma$ to the inverse of the DoS multiplied by $\pi$ $[\Gamma=(\mathrm{Im}G)^{-1}]$.
We start with Eq.~\eqref{sigma}. 
We note that the magnitude of the momentum $\bm{k}$ is $k_{\F}$ within the Andreev approximation. 
Hence, we omit the subscript of $\phi_{k_{\F}}$ and also use $\phi$ instead of $\bm{k}_{\F}$ as a label. 
We define $\bar{\phi}$, which satisfies $\bar{\phi} = 2\theta -\phi +\pi$; $\bar{s} = -s$ and $\bar{b} = b$. 
Without loss of generality, we can take $\phi$ such that $|\theta-\phi|\le \pi/2$ and, in this case, $s \ge 0$.
In order to use Eq.~\eqref{sigma}, it is necessary to construct an approximate eigenfunction for the BdG equation.
Here, we assume that we can approximately describe the eigenfunction using the following formula:
\begin{align}
\vec{u}_{l,\cc}(\bm{r}) = \dfrac{e^{\ii l\theta}}{\sqrt{2\pi N d\xi r}}\dfrac{\vec{u}_{\phi}(s,b)+ c_{2} \vec{u}_{\bar{\phi}}(s,b)}{2}, \label{wave-function}
\end{align}
where $d$ is the number of Fermi surfaces in the normal state and $c_{2}$ is the phase factor discussed in Appendix\ref{ap-s-wave}.
We can confirm that this formula is valid for an s-wave SC and a chiral p-wave SC through their analytical expression for the BdG wave function (we directly confirm this in Appendix\ref{ap-s-wave}). 
We need to obtain $\vec{u}_{\phi}$ from $\vec{\tilde{u}}^{\bb}_{\phi}$ 
through the unitary and gauge transformations 
since $\vec{u}_{\phi}$ used in Eq.~\eqref{wave-function} is described in the original basis. 
Moreover, we assume that the matrix element $m_{\phi_{1},\phi_{2}} \equiv \vec{\bar{u}}_{\phi_{1}}^{\dagger}\check{\tau}_{z}\vec{\bar{u}}_{\phi_{2}}$ 
satisfies the relation $m_{\phi_{1},\phi_{2}}= -m_{\phi_{2},\phi_{1}}$. This relation is satisfied in the present system, an s-wave SC, and a chiral p-wave SC. 
Under these assumptions, we evaluate the diagonal elements of Eq.~\eqref{mat-el}:
\begin{align}
M_{l,l} 
&= \dint r dr |2\pi \vec{u}_{l,c}^{\dagger}(\bm{r})\check{\tau}_{z}\vec{u}_{l,c}(\bm{r})|^{2} \nonumber \\
&= \dint  \dfrac{dr}{r} \dfrac{\left|m_{\phi,\phi}+m_{\bar{\phi},\bar{\phi}}+ \left(c_{2}m_{\phi,\bar{\phi}}e^{-2\ii k_{\F}s}+c_{2}^{*}m_{\bar{\phi},\phi}e^{2\ii k_{\F}s}\right)\right|^{2} }{(4N d \xi )^{2}} \nonumber \\
&= \dint  \dfrac{dr}{r} \dfrac{\left|m_{\phi,\bar{\phi}}\right|^{2} \left[1-\cos(4k_{\F}s-2\beta)\right]}{8(N d \xi )^{2}}. \label{matele}
\end{align}
In the last line, we put $c_{2} = e^{\ii \beta}$, and when $k_{\F}\xi$ is large, the integral of the second term is negligible.
This expression for the matrix element described by the Andreev wave function is general and we also calculate the scattering rates in the case of the s-wave SC using the formula in Appendix\ref{ap-s-wave}.
We evaluate the matrix element $|m_{\phi,\bar{\phi}}|^{2}$ as
\begin{align}
\left|m_{\phi,\bar{\phi}}\right| = &\left[ \dfrac{\sqrt{V_{z}^{2}+\alpha^{2}k_{\F}^{2}}-V_{z}}{\sqrt{V_{z}^{2}+\alpha^{2}k_{\F}^{2}}}\delta_{\kappa,-1}+\dfrac{\sqrt{V_{z}^{2}+\alpha^{2}k_{\F}^{2}}+V_{z}}{\sqrt{V_{z}^{2}+\alpha^{2}k_{\F}^{2}}}\delta_{\kappa,1}\right]\nonumber \\&\times\left|\sin(\phi-\bar{\phi})\right|\exp\left[-2\psi(s,b)\right]. \label{matele2}
\end{align}
This expression shows that the impurity effects are the same even if we change the sign of $V_{z}$ and $\kappa$ simultaneously. 
The difference between parallel and antiparallel vortices is described only by the coefficients $[(V_{z}^{2}+\alpha^{2}k_{F}^{2})^{1/2}\pm |V_{z}|]/(V_{z}^{2}+\alpha^{2}k_{F}^{2})^{1/2}$ [$+$ ($-$): parallel (antiparallel)].
In order to evaluate the integral, we use
\begin{align}
|\sin(\phi-\bar{\phi})| = 2|\sin(\theta-\phi)\cos(\theta-\phi)| = \dfrac{2|sb|}{s^{2}+b^{2}},
\end{align}
and perform the following approximation in the region of the integral. The integral in Eq.~\eqref{matele} can be evaluated as 
\begin{align}
\int_{b}^{\infty} \dfrac{dr}{r}\sin^{2}(\phi-\bar{\phi})e^{-4\psi(s,b)} 
&\simeq \int_{b}^{\xi}\dfrac{dr}{r}\dfrac{4s^{2}b^{2}}{(s^{2}+b^{2})^{4}} \nonumber \\
&=1-\dfrac{2b^{2}}{\xi^{2}}+\dfrac{b^{4}}{\xi^{4}}. \label{calc-1}
\end{align}
If we focus on the first excited state with angular momentum $|l| = 1$, we can estimate $b^{2}/\xi^{2} = 1/(k_{\F}\xi)^{2}$ and omit the second and third terms in Eq.~\eqref{calc-1}. We thus obtain the matrix element $M_{l,l}$ as 
\begin{align}
M_{l,l} =  \dfrac{1}{8(Nd \xi )^{2}}\left[ 1+\dfrac{\mathrm{sgn}(\kappa V_{z})\left|V_{z}\right|}{\sqrt{V_{z}^{2}+\alpha^{2}k_{\F}^{2}}}\right]^{2}.
\end{align}
We consider the two limiting cases of the scattering rates in detail. 
One is that there is a large level spacing in the presence of impurities and hence we can neglect the contribution from other angular momentum states. 
The other is the limit of continuous spectra, but the impurity strength is very small and thus we can treat it within 
the non-SCBA. 

In the former case, we neglect the contribution to $\sigma_{l}$ from other angular momentum states $l'$ in Eq.~\eqref{sigma}; hence, 
\begin{align}
\sigma_{l}(\ii\omega_{n}) &= \dfrac{\Gamma_{\mathrm{n}}}{2\pi^{2}N_{\mathrm{n}}}\dfrac{M_{l,l}}{E_{l,c}-\sigma_{l}(\ii\omega_{n})-\ii\omega_{n}},  \\
\sigma_{l}(E_{l,c}) &=\ii\sqrt{ \dfrac{\Gamma_{\mathrm{n}}}{2\pi^{2}N_{\mathrm{n}}}M_{l,l}},\\
\Gamma &\equiv\mathrm{Im}\sigma_{l}(E_{l,c})\nonumber \\
&=\Gamma_{\n}\sqrt{\dfrac{\pi\Delta_{\bb}}{8(\pi Nd)^{2}\Gamma_{\n}k_{\F}\xi}\left(1+\dfrac{\mathrm{sgn}(\kappa V_{z})\left|V_{z}\right|}{\sqrt{V_{z}^{2}+\alpha^{2}k_{\F}^{2}}}\right)^{2}}. \label{gamma-1}
\end{align}
In the second line, we perform the analytical continuation $\ii\omega_{n}\to \omega+\ii\delta$ and set $\omega=E_{l,c}$. We also use $N_{\n} = k_{\F}/(2\pi v_{\F})$ in the last line.

In the latter case, we replace the summation with respect to $l'$ by integration with respect to $b'$ through the relation $l' = -k_{\F} b'$. 
We focus on the low-energy states and thus use the approximate form of the energy spectrum 
$E_{l',c}= E(b')\simeq\kappa\Delta_{\mathrm{mini}} k_{\F}b' $ and put $\sigma_{l'}(\ii \omega_{n})=0$ in the denominator in Eq.~\eqref{sigma}, 
and then we can calculate $\Gamma$ as
\begin{align}
\Gamma &\equiv \mathrm{Im} \sigma_{l}(E(b)) = \dfrac{k_{\F}\Gamma_{\n}}{2\pi^{2}N_{\n}}\dint db'\mathrm{Im}\dfrac{M_{l(b)l(b')}}{E(b')-E(b)-\ii\delta} \nonumber\\
&= \dfrac{\Gamma_{\n} M_{l,l}}{2\pi N_{\n}\Delta_{\mathrm{mini}}} 
= \dfrac{\Gamma_{\n}}{8c_{1}(Nd)^{2}}\left(1+\dfrac{\mathrm{sgn}(\kappa V_{z})\left|V_{z}\right|}{\sqrt{V_{z}^{2}+\alpha^{2}k_{\F}^{2}}}\right)^{2}. \label{gamma-2}
\end{align}

\begin{figure}[t]
\begin{center}
\includegraphics[width = 7cm]{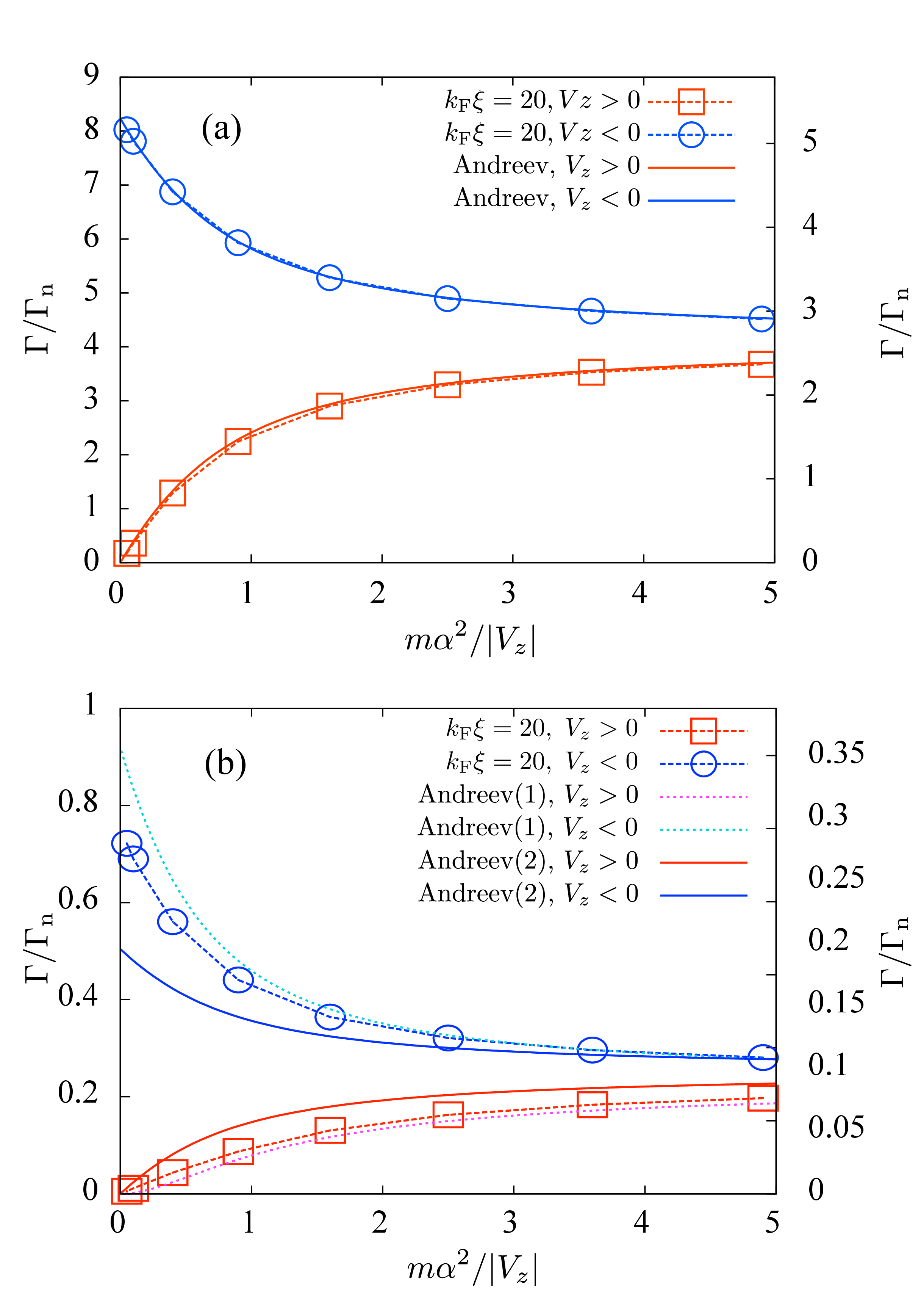}
\caption{(Color online) Scattering rates calculated by means of Andreev approximation [the solid lines labeled as ``Andreev(1)''  correspond to the calculation using Eq.~\eqref{gamma-1} and the dotted lines labeled as ``Andreev(2)'' correspond to that using Eq.~\eqref{gamma-2}] and 
the scheme represented by Eqs.~\eqref{green}~--~\eqref{mat-el} 
(open squares for $V_{z}>0$ and open circles for $V_{z}<0$). }
\label{g4g1}
\end{center}
\end{figure}
\begin{figure}[t]
\begin{center}
\includegraphics[width = 7cm]{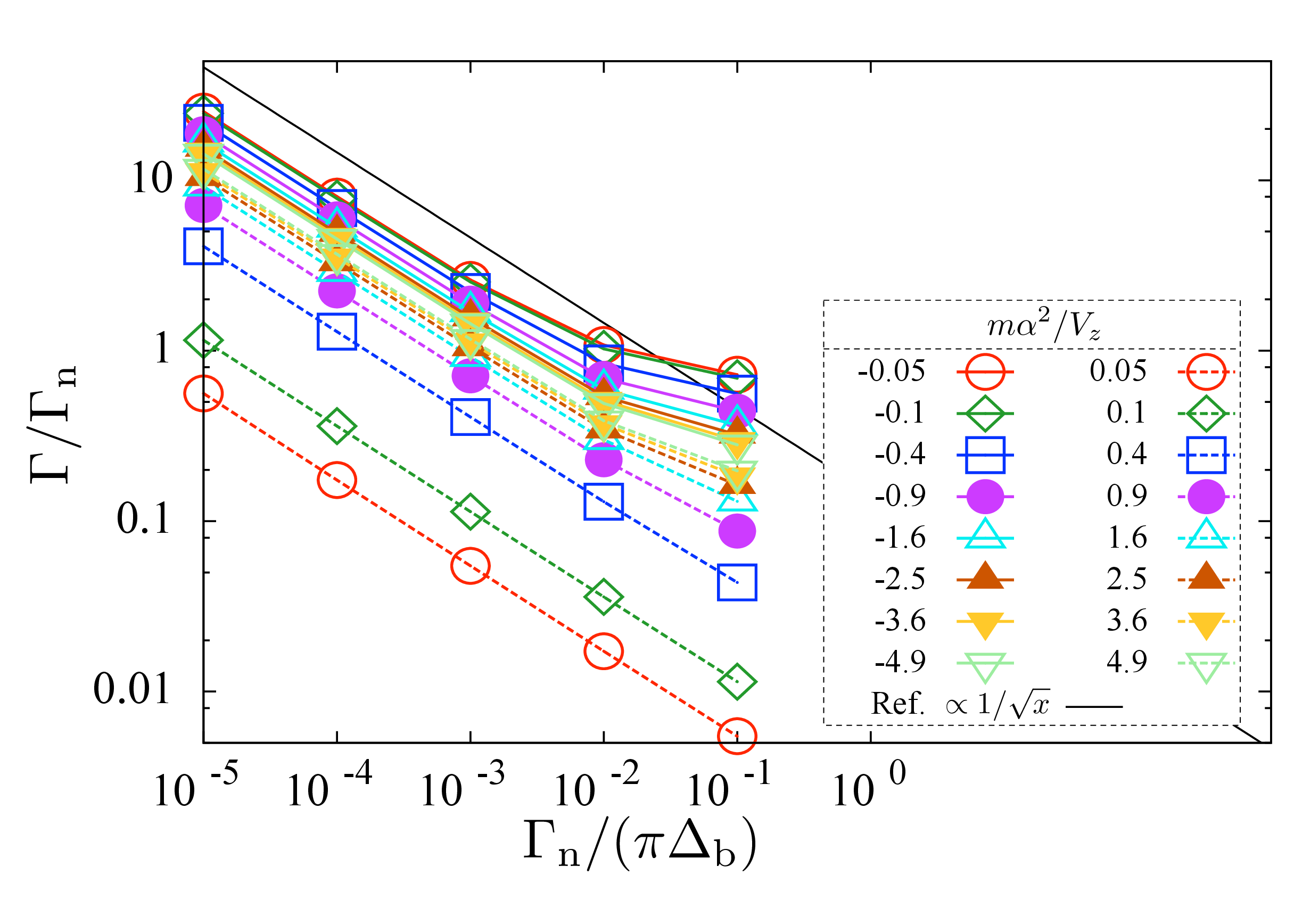}
\caption{(Color online) $\Gamma_{\n}$ dependence of scattering rate $\Gamma$. The solid line denotes the behavior of $[\Gamma_{\n}/(\pi\Delta_{\bb})]^{-1/2}$. }
\label{gvsgn}
\end{center}
\end{figure}

We compare the above two formulas with the numerical calculation. 
We again comment on the definition of scattering rates. 
In this section, in contrast to the previous section, we define $\Gamma$ as the inverse of $\mathrm{Im}G$ at the energy $E_{l,c}$. 
In Figs.~\ref{g4g1}(a) and \ref{g4g1}(b), we show the scattering rates for the first excited state with angular momentum $l=1$. 
The solid lines shown in Figs.~\ref{g4g1}(a) and \ref{g4g1}(b) represent the scattering rates calculated using Eq.~\eqref{gamma-1} 
while the dotted lines shown in Fig.~\ref{g4g1}(b) represent those calculated using Eq.~\eqref{gamma-2}. 
The dashed lines with open squares and open circles in Figs.~\ref{g4g1}(a) and \ref{g4g1}(b) represent the scattering rates based on the iKK scheme represented by Eqs.~\eqref{green}~--~\eqref{mat-el} for $V_{z}>0$ and $V_{z}<0$, respectively.  All the schemes show that when $V_{z}>0$, i.e., an antiparallel vortex, the scattering rate is an increasing function of $m\alpha^{2}/|V_{z}|$, and when $V_{z}<0$, i.e., a parallel vortex, 
the scattering rate is a decreasing function of $m\alpha^{2}/|V_{z}|$ as mentioned in Sect.~\ref{sec-num}.
We need to explain the vertical axis in Figs.~\ref{g4g1}(a) and \ref{g4g1}(b).
In Fig.~\ref{g4g1}(a), we set $\Gamma_{\n}/(\pi\Delta_{\bb}) = 10^{-4}$. 
The magnitude of the scattering rates based on the iKK scheme is shown on the left vertical axis and that based on Eq.~\eqref{gamma-1} is shown on the right vertical axis, 
the range of which is $0.64$ times that of the left axis. 
We tune the range in order to make the parameter dependence clear. 
In Fig.~\ref{g4g1}(b), we set $\Gamma_{\n} /(\pi\Delta_{\bb}) = 10^{-1}$. 
$\Gamma$ based on the iKK scheme is shown on the left axis and that based on Eq.~\eqref{gamma-1} is shown on the right axis, 
the range of which is $0.33$ times that of the left axis, as well as in Fig.~\ref{g4g1}(a). 
Moreover, $\Gamma$ calculated by multiplying Eq.~\eqref{gamma-2} by $\sqrt{2}$ is also shown on the left axis. 
(The aim of this operation is simply to make the parameter dependence clearer.)
The level spacing in this case is $\Delta_{\mathrm{mini}}/\Delta_{\bb} \sim (k_{\F}\xi)^{-1} = 0.05$, which is much larger than 
$\Gamma/\Delta_{\bb}\sim 4\pi \times10^{-4}$ in Fig.~\ref{g4g1}(a). 
In this case, we can consider that the system is in the limit of discrete spectra and that Eq.~\eqref{gamma-1} is valid. 
On the other hand, in Fig.~\ref{g4g1}(b), $\Gamma/\Delta_{\bb} \sim 0.2\pi \times 10^{-1}$, and this is comparable to the level spacing. 
Hence, Eq.~\eqref{gamma-1} no longer describes the $m\alpha^{2}/V_{z}$ dependence of $\Gamma/\Gamma_{\n}$ while this dependence is similar to 
Eq.~\eqref{gamma-2}, except in the region of very small $m\alpha^{2}/|V_{z}|$ for $V_{z}>0$. 
In this exceptional region, the system has discrete spectra, i.e., the level spacing $\Delta_{\mathrm{mini}}$ is always larger than the scattering rate $\Gamma$ 
since $\Gamma \to 0$ as $m\alpha^{2}/V_{z} \to 0$. 
We infer that Eq.~\eqref{gamma-2} does not explain the dependence precisely even when $V_{z}<0$ for the following two reasons. 
First, we cannot regard the system as being in the continuous limit because the scattering rate $\Gamma$ is not much larger than and comparable to the level spacing $\Delta_{\mathrm{mini}}$. 
Second, we should treat the impurities within the SCBA rather than the non-SCBA because the impurity strength $\Gamma_{\n}$ is large compared with the level spacing. 
We emphasize, however, that the $m\alpha^{2}/|V_{z}|$ dependence of $\Gamma/\Gamma_{\n}$ is similar to Eq.~\eqref{gamma-2} rather than Eq.~\eqref{gamma-1}.

Figure~\ref{gvsgn} shows the $\Gamma_{\mathrm{n}}/(\pi\Delta_{\bb})$ dependence of $\Gamma/\Gamma_{\n}$.
We expect $\Gamma/\Gamma_{\n}\propto 1/\sqrt{\Gamma_{\n}/(\pi\Delta_{\bb})}$ from Eq.~\eqref{gamma-1} and 
Fig.~\ref{gvsgn} implies that this is a good description for small $\Gamma_{\n}/(\pi\Delta_{\bb})$, 
approximately for $\Gamma_{\n}/(\pi\Delta_{\bb})$ such that $\Gamma k_{\F}\xi/(\pi\Delta_{\bb})\lesssim 0.1$. 

From the above discussion, we find that it is possible to describe impurity effects by the Andreev approximation. 
In the remaining part of this article, we discuss the origin of the scattering rates in terms of the ``spin-resolved angular momentum'' defined below 
and connect it to the ``total angular momentum'', which is the sum of the vorticity and chirality.
As mentioned in the text just below Eq.~\eqref{Hbdg} in Sect.~\ref{sec-model-method}, each component of the quasiparticle state labeled as $l$ has the angular momenta 
$(l-1\uparrow,l\downarrow,l+1\uparrow,l\downarrow)$ and $(l\uparrow,l+1\downarrow,l\uparrow,l-1\downarrow)$ in the original basis when $\kappa = -1$ and $1$, respectively. 
We can understand the origin of the scattering rates by dividing these four components into two pairs such that $(l\sigma,l\sigma)$ and $(l\pm1\sigma',l\mp1\sigma')$. 
We introduce the spin-resolved angular momentum to label these pairs: 
\begin{align}
L_{z\sigma} = -2 \delta_{\kappa,-1}\delta_{\sigma,\uparrow} + 2 \delta_{\kappa, 1}\delta_{\sigma,\downarrow}.
\end{align}
Here, we review the calculation of Eq.~\eqref{matele2} in terms of $L_{z\sigma}$.
We symbolically describe a part of the unitary transformation $\vec{u} \to \vec{u}^{\bb}$ as
\begin{align}
u^{\bb}_{-} = \sum_{\sigma} a_{-\sigma} u_{\sigma}, \ 
v^{\bb}_{-} = \sum_{\sigma} b_{-\sigma} v_{\sigma}.
\end{align}
We note that $|a_{-\sigma}| = |b_{-\sigma}|$ and $|a_{-\uparrow}|^{2}+|a_{-\downarrow}|^{2} =1$. The matrix element $m_{\phi\bar{\phi}}$ can also be described as 
\begin{align}
|m_{\phi\bar{\phi}}| 
\propto \left|\sum_{\sigma}|a_{-\sigma}|^{2}\sin\left(\dfrac{L_{z\sigma}}{2}(\phi-\bar{\phi})\right)\right| 
= |a_{-\tilde{\sigma}}|^{2}\left|\sin\left(\phi-\bar{\phi}\right)\right|.
\end{align}
$\tilde{\sigma}$ denotes the spin component with $|L_{z\tilde{\sigma}}| = 2$ 
while the other spin component with $L_{z\sigma} = 0$ does not contribute to the scattering rates. 
We remark that $|a_{-\sigma}|(\sigma = \uparrow,\downarrow)$ are the degrees of mixing of $u_{\uparrow}$ and $u_{\downarrow}$ to make the Fermi surface.
Therefore, the scattering rates are determined by the ratio of the contribution from the spin component with $|L_{z\sigma}| = 2$ to the Fermi surface.

Finally, we introduce the total angular momentum $\check{L}_{z} = -\ii 2\check{\tau}_{z}(\partial_{\theta} +\partial_{\phi})$, 
which acts on the wave function $e^{-\ii l\theta} \vec{u}_{l,c} (\bm{r})$. 
We can corroborate that this definition is consistent
with Eq.~(2.9) in Ref.~\citen{KatoHayashi2001} for the s-wave SC and chiral p-wave SC.
We calculate the expectation value 
\begin{align}
\braket{L_{z}} 
&= \Tr \left[ \vec{u}_{l,c}^{\dagger}(\bm{r}) e^{\ii l \theta}\check{L}_{z}e^{-\ii l\theta} \vec{u}_{l,c} (\bm{r})\right] \nonumber \\
&=\sum_{\sigma} |a_{-\sigma}|^{2}L_{z\sigma} = |a_{-\tilde{\sigma}}|^{2} L_{z\tilde{\sigma}}.
\end{align}
Therefore, we can understand the impurity scattering rates as the magnitude of the total angular momentum. 
Both limiting cases, $|\braket{L_{z}}|= 0$ and $2$, where $L_{z}$ is a good quantum number, correspond to chiral p-wave SCs. 
On the other hand, it is remarkable that $|\braket{L_{z}}| = 1$ does not describe the s-wave SC because $L_{z}$ is no longer a good quantum number.
$|\braket{L_{z}}| = 1$ is simply due to the superposition of $|\braket{L_{z}}|=0$ and $|\braket{L_{z}}|=2$

\section{Summary}\label{summary}
In this paper, we have investigated the impurity effects on the bound (CdGM) states in a vortex core of a topological s-wave SC.
We have calculated the scattering rates for CdGM states numerically and analytically. 
The numerical calculation is performed for discrete spectra based on the iKK scheme in order to discuss the impurity effects on the level spacing of bound states.
We have found that there are similarities and differences compared with the chiral p-wave SC: 
the parameter $m\alpha^{2}/V_{z}$ is important for describing the impurity effects as well as the relative sign of the vorticity and chirality. 
On the other hand, the analytical calculation is based on the Andreev approximation to understand the physical picture of impurity scattering and the origin of the parameter-dependent scattering rates.
The conclusion is that the scattering rates $|a_{-\sigma}|^{2}$ are determined 
by the ratio of the spin components with $|L_{z\sigma}|=2$ composing the Fermi surface, 
and this ratio describes the magnitude of the total angular momentum $|\braket{L_{z}}| = 2 |a_{-\sigma}|^{2}$. 

Although the topological s-wave SC belongs to the same class as the chiral p-wave SC, the properties of excited states are more complicated than those in the chiral p-wave SC. 
However, vanishingly small scattering rates because of the coherence factors and rotational symmetry are obtained as well as in the case of the chiral p-wave SC. 
The topological s-wave SC is more promising than a chiral p-wave SC such as Sr$_{2}$RuO$_{4}$ 
in terms of the robust low-energy states in a vortex core in TSC owing to the continuous rotational symmetry 
since this superconductivity is expected to appear in the 2DEG in the semiconductor heterostructure 
while the chiral p-wave superconductivity in Sr$_{2}$RuO$_{4}$ is affected by the $C_{4}$ point group symmetry.

\section*{Acknowledgements}

This work was supported by the Program for Leading Graduate 
Schools, MEXT, Japan and by KAKENHI(No. 23244070) from JSPS.

\appendix
\section{Edge state} \label{ap-edge}
We construct the zero-energy edge state, which appears near the edge of a TSC, in order to confirm that the Andreev approximation can describe the features of a TSC. 
For simplicity, we consider the one-dimensional system of a topological s-wave SC in the region $x\ge0$ and vacuum in the region $x<0$. 
The BdG equation for the homogeneous topological s-wave SC in one dimension is as follows:
\begin{align}
\check{H}(k) \vec{\bar{u}}_{k} &= E \vec{\bar{u}}_{k}, \\
\check{H}(k) &=\epsilon_{k} \check{\tau}_{z} + V_{z} \check{\sigma}_{z}\check{\tau}_{z} -\alpha k \check{\sigma}_{y}\check{\tau}_{z} -\Delta\check{\sigma}_{y}\check{\tau}_{y},
\end{align}
where $k$ is the one-dimensional momentum. $\check{\sigma}$ and $\check{\tau}$ denote Pauli matrices in spin space and Nambu space, respectively, and they are 4 by 4 matrices.
The vector $\vec{\bar{u}}_{k}$ has four components, 
$\vec{\bar{u}}_{k} = \trans(u_{k\uparrow},u_{k\downarrow},v_{k\uparrow},v_{k\downarrow})$. 
In a similar way to Sect.~\ref{sec-andreev}, we can write down the Andreev equation by introducing a slow spatial variance as an edge at $x=0$, but in this case, we impose the boundary condition where the wave function $\vec{u}$ is $0$. 
Our goal is to construct the low-energy, particularly the zero-energy wave function satisfying the boundary condition
by linear combination of the zero-energy solutions to the Andreev equation. 
The Andreev equation with $k=\pm k_{\F}$ can be written as 
\begin{align}
\mp\left[\ii v_{\F} \partial_{x}\hat{\sigma}_{z} +f_{p}\Delta(x) \hat{\sigma}_{y}\right]\vec{\underline{\bar{u}}}_{\pm k_{\F}-}^{\bb} =0.
\end{align}
The solutions to these equations are given by
\begin{align}
\vec{\underline{u}}_{\pm k_{\F}-}^{\bb} = 
\begin{pmatrix}
1 \\ -1 
\end{pmatrix}
\exp\left[\ii k_{\F}x-\dfrac{1}{v_{\F}}\int_{0}^{x}f_{p}\Delta(x') dx'\right].
\end{align}
The unitary matrix $\check{U}_{k}$ defined as $\vec{\bar{u}}_{k} = \check{U}_{k}\vec{\bar{u}}_{k}^{\bb}$ is given by
\begin{align}
\check{U}_{k} = \left[\ii\alpha k\check{1}+\left(\sqrt{V_{z}^{2}+\alpha^{2}k^{2}}-V_{z}\right)\check{\sigma}_{x}\right]/C,
\end{align}
where $C$ is defined by Eq.~\eqref{u-norm}.
In addition to these solutions with Fermi momenta, we construct the zero-energy solutions with $k\sim 0$.
In this case, we can write down the Andreev equation as
\begin{align}
\left[V_{z}\check{\sigma}_{z}\check{\tau}_{z} -\mu \check{\tau}_{z} +\ii \alpha \partial_{x} \check{\sigma}_{y}\check{\tau}_{z}
- \Delta\check{\sigma}_{y}\check{\tau}_{y} \right] \vec{u}_{0} = 0.
\end{align}
This eigenvectors can be obtained by considering the matrix
\begin{align}
\check{B} = \left[V_{z}\check{\sigma}_{x}+\ii\mu \check{\sigma}_{y}+\Delta\check{\tau}_{x}\right]/\alpha.
\end{align}
We can construct the zero-energy solutions with positive eigenvalues $\lambda_{i} (i=1,2,\cdots)$ 
and the corresponding eigenvectors $\vec{b}_{i}$ of $\check{B}$. 
When $V_{z}^{2}>\Delta^{2}+\mu^{2}$, there are two positive eigenvalues and eigenvectors: 
\begin{align}
\alpha \lambda_{1} &= \sqrt{V_{z}^{2}-\mu^{2}}-\Delta,\  \vec{b}_{1} = \trans(-b_{+},-b_{-},b_{+},b_{-}),\\
\alpha \lambda_{2} &= \sqrt{V_{z}^{2}-\mu^{2}}-\Delta,\  \vec{b}_{2} = \trans(b_{+},b_{-},b_{+},b_{-}),\\
b_{+} &= \sqrt{V_{z}+\mu}/2|V_{z}|,\  b_{-} = \sqrt{V_{z}-\mu}/2|V_{z}|.
\end{align}
With these sets of eigenvalues and eigenvectors, wave functions can be constructed as
\begin{align}
\vec{u}_{0\lambda_{i}} = \vec{b}_{i} \exp\left[-\int_{0}^{x} \lambda_{i} dx'\right] \ (i=1,2),
\end{align}
which decay far from the edge at $x=0$. We consider the linear combination of these bases:  
\begin{align}
\vec{u} (x) = C_{+}\vec{u}_{+k_{\F}} + C_{-} \vec{u}_{-k_{\F}}+C_{1}\vec{u}_{0\lambda_{1}} + C_{2} \vec{u}_{0\lambda_{2}}. \label{eq-lc}
\end{align}
Here, we omit the contribution from $\vec{\underline{u}}_{\pm k_{\F}+}^{\bb}$ and perform the unitary transformation $\check{U}_{\pm k_{\F}}$ to obtain $\vec{u}_{\pm k_{\F}}$.
Since we can choose $C_{+}$ and $C_{-}$ such that $C_{+}\vec{u}_{+k_{\F}} + C_{-} \vec{u}_{-k_{\F}}$ is parallel to $\vec{u}_{0\lambda_{1}}$ at $x=0$, 
we find that $C_{2}=0$. We emphasize that it is necessary to consider the $k=0$ components only for the edge state rather than the vortex core state.

\section{Scattering rates for s-wave SC}\label{ap-s-wave}
The BdG wave function for the CdGM states in a vortex core for the s-wave or chiral p-wave SC and its asymptotic behavior for large $k_{\F}r$ are described as 
\begin{align}
\vec{\underline{u}}_{l,c}&=\begin{pmatrix}
u_{l,c\uparrow}(\bm{r}) \\v_{l,c\downarrow}(\bm{r})
\end{pmatrix}
=\dfrac{1}{\sqrt{\tilde{N}}}\begin{pmatrix}
J_{l-L_{z}/2} (k_{\F}r)e^{-K(r)} e^{\ii (l-L_{z}/2)\theta} \\
-\ii e^{\ii L_{z}\pi /2}J_{l+L_{z}/2} (k_{\F}r) e^{-K(r)}e^{\ii (l+L_{z}/2)\theta}
\end{pmatrix}\nonumber \\
&\sim\dfrac{e^{\ii l\theta-K(r)}}{\sqrt{2\pi \tilde{N}k_{\F}r}}\begin{pmatrix}
e^{-\ii \theta/2}\left[e^{\ii g_{l-L_{z}/2}(k_{\F}r)}+e^{-\ii g_{l-L_{z}/2}(k_{\F}r)}\right]  \\
-\ii e^{\ii L_{z}\pi /2 + \ii \theta/2}\left[e^{\ii g_{l+L_{z}/2}(k_{\F}r)}+e^{-\ii g_{l+L_{z}/2}(k_{\F}r)}\right]
\end{pmatrix},\label{bdg-expand1}\\
K(r) &= \dfrac{1}{v_{\F}}\int_{0}^{r}\Delta(r') dr', \\
g_{l}(x) &= x+ l^{2}/(2x)-(2l+1)\pi/4,
\end{align}
where $\tilde{N}$ is a normalization constant. 
$-L_{z}$ is the sum of the vorticity and chirality and for example, $L_{z}$ takes $\mp 1$ for the s-wave SC, 
$\mp2$ for the chiral p-wave SC with the parallel vortex, and $0$ for the chiral p-wave SC with the antiparallel vortex. 
For convenience, we adopt the minus sign in the definition of $L_{z}$.
Equation~\eqref{bdg-expand1} is reduced to 
\begin{align}
&\begin{pmatrix}
u_{l,c\uparrow}(\bm{r}) \\v_{l,c\downarrow}(\bm{r})
\end{pmatrix}
\sim\dfrac{c_{3}e^{\ii l\theta-K(r)}}{\sqrt{2\pi \tilde{N}k_{\F}r}}\nonumber \\
&\times\begin{pmatrix}
e^{-\ii \theta/2}\left[e^{\ii [g_{1,l}(k_{\F}r)-g_{2,l}(k_{\F}r)]}+\ii c_{4}e^{-\ii [g_{1,l}(k_{\F}r)-g_{2,l}(k_{\F}r)]}\right]  \\
-\ii e^{ \ii \theta/2}\left[e^{\ii [g_{1,l}(k_{\F}r)+g_{2,l}(k_{\F}r)]}+\ii c_{4}e^{-\ii [g_{1,l}(k_{\F}r)+g_{2,l}(k_{\F}r)]}\right]
\end{pmatrix},\\
&g_{l,1}(x) = x +\left(l^{2}+L_{z}^{2}/4\right)/(2x), \\
&g_{l,2}(x) =L_{z}l/(2x), \\
&c_{3} = e^{-\ii (2l-L_{z}+1)\pi/4}, \\
&c_{4} = e^{\ii (2l-L_{z})\pi/2}.
\end{align}
$c_{3}$ is simply an overall phase factor and we need not consider it. 
We take an approximation that $g_{l,1}(k_{\F}r) \sim k_{\F}|s|$. 
Of course, this approximation is not valid in the region where $k_{\F}|s|$ is not large compared with $|l|=k_{\F}|b|$, 
but we expect that the physical picture does not change significantly.

On the other hand, the Andreev wave function is described as
\begin{align}
\begin{pmatrix}
u_{\phi} \\
v_{\phi}
\end{pmatrix}
&=\begin{pmatrix}
e^{-\ii L_{z}\phi/2}\\
e^{\ii L_{z}\phi/2}
\end{pmatrix}
\exp[\ii k_{\F} s-\psi(s,b)], \\
\psi(s,b) &= \dfrac{1}{v_{\F}}\int_{0}^{s}\Delta(r')\dfrac{s'}{r'} ds'.
\end{align}
As mentioned in the main text, we take $\phi$ such that $|\theta-\phi|<\pi/2$, hence, $s\ge 0$. 
We consider a linear combination of two momenta $\phi$ and $\bar{\phi}$, which share the impact parameter $b$,
\begin{align}
&e^{\psi(s,b)}[\vec{\underline{u}}_{\phi} + c_{2} \vec{\underline{u}}_{\bar{\phi}}] \nonumber\\
&\simeq
\begin{pmatrix}
[e^{\ii k_{\F}s+\ii L_{z}b/(2r)} + c_{2} e^{-\ii L_{z} \pi /2}e^{-\ii k_{\F}s-\ii L_{z}b/(2r)} ]e^{-\ii L_{z}\theta /2} \\
-\ii[e^{\ii k_{\F}-\ii L_{z}b/(2r)} + c_{2} e^{\ii L_{z} \pi /2}e^{-\ii k_{\F}s+\ii L_{z}b/(2r)}] e^{\ii L_{z}\theta /2}
\end{pmatrix}\nonumber\\
&=
\begin{pmatrix}
[e^{\ii k_{\F}s-\ii L_{z}l/(2k_{\F}r)} + c_{2} e^{-\ii L_{z} \pi /2}e^{-\ii k_{\F}s+\ii L_{z}l/(2k_{\F}r)} ]e^{-\ii L_{z}\theta /2} \\
-\ii[e^{\ii k_{\F}s+\ii L_{z}l/(2k_{\F}r)} + c_{2} e^{\ii L_{z} \pi /2}e^{-\ii k_{\F}s-\ii L_{z}l/(2k_{\F}r)}] e^{\ii L_{z}\theta /2}
\end{pmatrix},
\end{align}
where we use $\theta-\phi\simeq \sin(\theta-\phi) =b/r$ in the second line and $l =-k_{\F}b$ in the third line.
When $b$ is small compared with $s$,  $K(r) \sim \psi(s,b)$, and thus we obtain
\begin{align}
c_{2} = \ii c_{4} = e^{\ii(l+1/2)\pi}, \\
\vec{\underline{u}}_{l,c} = \dfrac{e^{\ii l\theta}}{\sqrt{2\pi\tilde{N} k_{\F} r}}\left[\vec{\underline{u}}_{\phi} + c_{2} \vec{\underline{u}}_{\bar{\phi}}\right].
\end{align} 
We can choose the normalization constant $\tilde{N} = 4N\xi d/k_{\F}$, where $N$ is defined by Eq.~\eqref{Norm} and $d$ is the number of Fermi surfaces; in this case, $d=1$.
Equation~\eqref{wave-function} can thus be obtained in the case of s-wave and chiral p-wave SCs.

In the remaining part of this appendix, using Eqs.~\eqref{matele} and \eqref{gamma-2}, we calculate the scattering rates of the CdGM mode in a vortex core in an s-wave SC. In this case, 
$l$ is a half-odd-integer and $L_{z} = \pm1$. We evaluate the matrix element $M_{l,l}$ using Eq.~\eqref{matele},
\begin{align}
M_{l,l} 
&= \int \dfrac{dr}{r} \dfrac{\left|m_{\phi\bar{\phi}}\right|^{2}}{8N^{2}\xi^{2}} \nonumber \\
&= \int_{b}^{\infty} \dfrac{dr}{r} \dfrac{\left|2\ii \sin\left[L_{z}(\phi-\bar{\phi})/2\right]\right|^{2}}{8N^{2}\xi^{2}}e^{-4\psi(s,b)}\nonumber \\
&\simeq \int_{0}^{\xi} ds\dfrac{s}{r^{2}} \dfrac{\cos^{2}(\theta-\phi)}{2N^{2}\xi^{2}} \nonumber\\
&= \int_{0}^{\xi} ds\dfrac{s^{3}}{r^{4}} \dfrac{1}{2N^{2}\xi^{2}} \nonumber \\
&= -\dfrac{1}{2N^{2}\xi^{2}}\left\{\ln |b| +\dfrac{1}{2} \left[1-\ln (\xi^{2}+b^{2})-\dfrac{b^{2}}{\xi^{2}+b^{2}}\right]\right\} \nonumber \\
&\to-\dfrac{1}{2N^{2}\xi^{2}}\ln (|b|/\xi) \quad (\text{as}\  b\to 0).
\end{align}
Noting that $|E(b)|= \Delta_{\mathrm{mini}}k_{\F}|b|$, the scattering rate is obtained as
\begin{align}
\dfrac{\Gamma}{\Gamma_{\n}} 
&= -\dfrac{1}{4\pi N^{2}\xi^{2} N_{\n}\Delta_{\mathrm{mini}}} \ln \left[\dfrac{E(b)}{\Delta_{\mathrm{mini}}k_{\F}\xi}\right] \nonumber \\
&=-\dfrac{\pi}{c_{1}N^{2}}\ln \left[\dfrac{E(b)}{\Delta_{\mathrm{mini}}k_{\F}\xi}\right],
\end{align}
where the coefficient is on the order of 1. This logarithmic dependence of $\Gamma$ on $E(b)$ is in good agreement with earlier works based on the quasiclassical theory~\cite{kopnin1994sign,PhysRevB.51.15291}. 
Similarly, we can calculate the scattering rates of CdGM states for chiral p-wave vortices.

\bibliographystyle{jpsj}
\bibliography{reference}

\begin{thebibliography}{10}

\bibitem{PhysRevLett.86.268}
D.~A. Ivanov: Phys. Rev. Lett. {\bfseries 86} (2001) 268.

\bibitem{RevModPhys.80.1083}
C.~Nayak, S.~H. Simon, A.~Stern, M.~Freedman, and S.~Das~Sarma: Rev. Mod. Phys.
  {\bfseries 80} (2008) 1083.

\bibitem{RevModPhys.82.3045}
M.~Z. Hasan and C.~L. Kane: Rev. Mod. Phys. {\bfseries 82} (2010) 3045.

\bibitem{RevModPhys.83.1057}
X.-L. Qi and S.-C. Zhang: Rev. Mod. Phys. {\bfseries 83} (2011) 1057.

\bibitem{alicea2011non}
J.~Alicea, Y.~Oreg, G.~Refael, F.~von Oppen, and M.~P.~A. Fisher: Nat. Phys.
  {\bfseries 7} (2011) 412.

\bibitem{PhysRevB.82.174506}
L.~Mao and C.~Zhang: Phys. Rev. B {\bfseries 82} (2010) 174506.

\bibitem{PhysRevB.82.020509}
A.~R. Akhmerov: Phys. Rev. B {\bfseries 82} (2010) 020509.

\bibitem{Caroli1964}
C.~Caroli, P.~G. de~Gennes, and J.~Matricon: Phys. Lett. {\bfseries 9} (1964)
  307 .

\bibitem{RevModPhys.59.533}
M.~M. Salomaa and G.~E. Volovik: Rev. Mod. Phys. {\bfseries 59} (1987) 533.

\bibitem{volovik1999fermion}
G.~E. Volovik: JETP Lett. {\bfseries 70} (1999) 609.

\bibitem{PhysRevB.61.10267}
N.~Read and D.~Green: Phys. Rev. B {\bfseries 61} (2000) 10267.

\bibitem{PhysRevB.55.1142}
A.~Altland and M.~R. Zirnbauer: Phys. Rev. B {\bfseries 55} (1997) 1142.

\bibitem{PhysRevB.78.195125}
A.~P. Schnyder, S.~Ryu, A.~Furusaki, and A.~W.~W. Ludwig: Phys. Rev. B
  {\bfseries 78} (2008) 195125.

\bibitem{ryu2010topological}
S.~Ryu, A.~P. Schnyder, A.~Furusaki, and A.~W.~W. Ludwig: New J. Phys.
  {\bfseries 12} (2010) 065010.

\bibitem{Kato2000}
Y.~Kato: J. Phys. Soc. Jpn. {\bfseries 69} (2000) 3378.

\bibitem{JPSJ.71.1721}
Y.~Kato and N.~Hayashi: J. Phys. Soc. Jpn. {\bfseries 71} (2002) 1721.

\bibitem{MaenoKittakaNomuraYonezawaIshida2012}
Y.~Maeno, S.~Kittaka, T.~Nomura, S.~Yonezawa, and K.~Ishida: J. Phys. Soc. Jpn.
  {\bfseries 81} (2012) 011009.

\bibitem{Moore1991}
G.~Moore and N.~Read: Nucl. Phys. B {\bfseries 360} (1991) 362 .

\bibitem{PhysRevLett.100.096407}
L.~Fu and C.~L. Kane: Phys. Rev. Lett. {\bfseries 100} (2008) 096407.

\bibitem{PhysRevLett.104.040502}
J.~D. Sau, R.~M. Lutchyn, S.~Tewari, and S.~Das~Sarma: Phys. Rev. Lett.
  {\bfseries 104} (2010) 040502.

\bibitem{Eilenberger1968}
G.~Eilenberger: Z. Phys. {\bfseries 214} (1968) 195.

\bibitem{larkin1969quasiclassical}
A.~I. Larkin and Y.~N. Ovchinnikov: Sov. Phys. JETP {\bfseries 28} (1969) 1200.

\bibitem{RevModPhys.78.373}
A.~V. Balatsky, I.~Vekhter, and J.-X. Zhu: Rev. Mod. Phys. {\bfseries 78}
  (2006) 373.

\bibitem{PhysRevB.57.5457}
A.~I. Larkin and Y.~N. Ovchinnikov: Phys. Rev. B {\bfseries 57} (1998) 5457.

\bibitem{PhysRevB.59.12021}
A.~A. Koulakov and A.~I. Larkin: Phys. Rev. B {\bfseries 59} (1999) 12021.

\bibitem{PhysRevB.60.14597}
A.~A. Koulakov and A.~I. Larkin: Phys. Rev. B {\bfseries 60} (1999) 14597.

\bibitem{kopnin1976conductivity}
N.~B. Kopnin and V.~E. Kravtsov: JETP Lett. {\bfseries 23} (1976) 578.

\bibitem{masakikato}
Y.~Masaki and Y.~Kato: arXiv:1505.02553 (2015).

\bibitem{Wu2012}
L.-H. Wu, Q.-F. Liang, Z.~Wang, and X.~Hu: J. Phys.: Conf. Ser. {\bfseries 393}
  (2012) 012018.

\bibitem{andreev1964thermal}
A.~F. Andreev: Sov. Phys. JETP {\bfseries 19} (1964) 1228.

\bibitem{volovik1993vortex}
G.~E. Volovik: JETP Lett. {\bfseries 57} (1993) 244.

\bibitem{Tewari2010219}
S.~Tewari, J.~D. Sau, and S.~D. Sarma: Ann. Phys. {\bfseries 325} (2010) 219 .

\bibitem{PhysRevB.81.125318}
J.~Alicea: Phys. Rev. B {\bfseries 81} (2010) 125318.

\bibitem{PhysRevD.13.3398}
R.~Jackiw and C.~Rebbi: Phys. Rev. D {\bfseries 13} (1976) 3398.

\bibitem{KatoHayashi2001}
Y.~Kato and N.~Hayashi: J. Phys. Soc. Jpn. {\bfseries 70} (2001) 3368.

\bibitem{kopnin1994sign}
N.~B. Kopnin: JETP Lett. {\bfseries 60} (1994) 130.

\bibitem{PhysRevB.51.15291}
N.~B. Kopnin and A.~V. Lopatin: Phys. Rev. B {\bfseries 51} (1995) 15291.

\end{thebibliography}


\end{document}